\documentclass[12pt]{article}
\usepackage{graphicx,anysize,amsfonts,amsmath,amsthm,amssymb,enumerate,hyperref,color,setspace,booktabs,enumitem, subcaption, mathtools}
\usepackage[font = normalsize]{caption}
\usepackage[mathscr]{euscript}
\usepackage[utf8]{inputenc}
\usepackage[english]{babel}
\usepackage{natbib}
\usepackage{moreverb, ragged2e}
\usepackage[font = normalsize]{subcaption}
\usepackage{rotating}
\newcommand\BibTeX{{\rmfamily B\kern-.05em \textsc{i\kern-.025em b}\kern-.08em
T\kern-.1667em\lower.7ex\hbox{E}\kern-.125emX}}
\newcommand\numberthis{\addtocounter{equation}{1}\tag{\theequation}}

\DeclarePairedDelimiter\ceil{\lceil}{\rceil}

\allowdisplaybreaks
\marginsize{1in}{1in}{1in}{1in}

\usepackage{sectsty}
\sectionfont{\fontsize{12}{15}\selectfont}
\subsectionfont{\fontsize{12}{15}\selectfont}

\usepackage{array}
\newcolumntype{L}[1]{>{\raggedright\let\newline\\\arraybackslash\hspace{0pt}}m{#1}}
\newcolumntype{C}[1]{>{\centering\let\newline\\\arraybackslash\hspace{0pt}}m{#1}}
\newcolumntype{R}[1]{>{\raggedleft\let\newline\\\arraybackslash\hspace{0pt}}m{#1}}
\providecommand{\keywords}[1]
{
  \small	
  \textbf{\textit{Keywords:}} #1
}

\hypersetup{hidelinks}
\title{\Large \textbf{Unified and Simple Sample Size Calculations for Individual or Cluster Randomized Trials with Skewed or Ordinal Outcomes}}
\author{\vspace{-1em} \small Shengxin Tu$^{1}$,  Chun Li$^{5}$, Caroline De Schacht$^{6}$, Carolyn M. Audet$^{2,4}$, \\ \small Aminu Taura Abdullahi$^{7}$, Edwin Trevathan$^{3,4}$, and Bryan E. Shepherd$^{1,4*}$ \\\vspace{-1em} \small
$^{1}$Department of Biostatistics, $^{2}$Department of Health Policy, \\\vspace{-1em} \small $^{3}$Departments of Pediatrics and Neurology,\\\vspace{-1em} \small $^{4}$Vanderbilt Institute for Global Health, Vanderbilt University Medical Center,\\\vspace{-1em} \small Nashville, Tennessee, USA.\\\vspace{-1em} \small
$^{5}$Department of Population and Public Health Sciences, University of Southern California,\\\vspace{-1em} \small Los Angeles, California, USA.\\\vspace{-1em} \small
$^{6}$ Friends in Global Health, Maputo, Mozambique\\\vspace{-1em} \small
$^{7}$Department of Psychiatry, Aminu Kano Teaching Hospital, Bayero University Kano, \\\vspace{-1em} \small Kano, Nigeria. \\
\small $^{*}$bryan.shepherd@vanderbilt.edu}
\date{\vspace{-2ex}}

\begin{document}
\maketitle
\begin{abstract}
Sample size calculations can be challenging with skewed continuous outcomes in randomized controlled trials (RCTs). Standard t-test-based calculations may require data transformation, which may be difficult before data collection. Calculations based on individual and clustered Wilcoxon rank-sum tests have been proposed as alternatives, but these calculations assume no ties in continuous outcomes, and clustered Wilcoxon rank-sum tests perform poorly with heterogeneous cluster sizes. Recent work has shown that continuous outcomes can be analyzed in a robust manner using ordinal cumulative probability models. Analogously, sample size calculations for ordinal outcomes can be applied as a robust design strategy for continuous outcomes. We show that Whitehead’s sample size calculations for independent ordinal outcomes can be easily extended to continuous outcomes. We extend these calculations to cluster RCTs using a design effect incorporating the rank intraclass correlation coefficient. Therefore, we provide a unifying and simple approach for designing individual and cluster RCTs that makes minimal assumptions on the distribution of the still-to-be-collected outcome. We conduct simulations to evaluate our approach’s performance and illustrate its application in multiple RCTs: an individual RCT with skewed continuous outcomes, a cluster RCT with skewed continuous outcomes, and a non-inferiority cluster RCT with an irregularly distributed count outcome.
\end{abstract}
\clearpage

\pagenumbering{arabic} 

\section{Introduction}
\label{s:intro}

The outcome of interest in randomized controlled trials (RCTs) is often a skewed continuous variable. For example, the Homens para $\text{Sa}\acute{\text{u}}\text{de}$ Mais (HoPS+) study conducted a cluster RCT to measure the impact of a multi-component intervention on adherence to antiretroviral therapy (ART) of pregnant women living with HIV \citep{audet2018, audet2024}. The primary outcome was adherence to treatment (i.e., the proportion of medications taken within 1 year), which is pseudo-continuous ranging from 0 to 1 with a left-skewed distribution.

Sample size calculations can be challenging with skewed continuous response variables. Standard sample size calculations based on the t-test may not apply unless data are transformed. But choosing an appropriate transformation and deciding an effect size on the transformed scale may be difficult, particularly prior to actually collecting data. These challenges are compounded in cluster RCTs, where observations from individuals within the same cluster may tend to be more similar than observations from different clusters, introducing complexity to their designs.

To avoid the challenges of transforming skewed continuous outcomes in RCTs, one can fit rank-based methods such as the Wilcoxon rank sum test, and one can perform sample size calculations based on these tests. Power and sample size calculations based on the Wilcoxon rank sum test exist \citep{rosner2009}. In addition, these calculations have been extended for the clustered Wilcoxon rank-sum test \citep{rosner2003, rosner2011}, which can be used for the design of cluster RCTs. However, these calculations are complicated; they assume the outcome is continuous with no ties, and the clustered Wilcoxon rank-sum test has problems when there are heterogeneous cluster sizes as we will illustrate. 

Recent work has demonstrated that continuous outcomes can be analyzed with ordinal cumulative probability models (CPMs) \citep{liu2017, mccullagh1980}. In these models, each unique level of the continuous outcome is treated as its own ordinal category; in the absence of ties, the number of ordinal categories is equal to the sample size. Fitting a CPM to the continuous response data is equivalent to fitting a semiparametric linear transformation model \citep{zeng2007,liu2017}, where the transformation is left unspecified and estimated as part of the model fitting procedure. Thus, CPMs are particularly useful with skewed or mixed-type outcome variables \citep{liu2017, tian2024addressing}. CPMs are rank-based, and the test statistic for a binary treatment effect from a CPM is closely related to the Wilcoxon rank sum test \citep{whitehead1993}. In addition, CPMs have been extended to allow the use of these models with clustered data \citep{tian2023}.

Just as ordinal cumulative probability models can be applied as a robust analysis approach for continuous response variables, sample size/power calculations for ordinal variables can be applied as a robust strategy for the design of RCTs with continuous response variables. Simple sample size calculations for independent ordinal response variables were derived by Whitehead \citep{whitehead1993}. In this manuscript, we show that these calculations are applicable not only to ordinal response variables but also to continuous response variables. Thus, they provide a unifying and simple approach for designing RCTs that makes minimal assumptions on the distribution of the still-to-be-collected response variable. 

We will also show that the calculations of Whitehead \citep{whitehead1993} can be easily extended to compute sample sizes for cluster RCTs with continuous or ordinal outcomes. A conventional and simple approach to calculating sample sizes in cluster RCTs is to inflate the sample size of an adequately powered individual RCT by the design effect (DE) based on the intraclass correlation coefficient (ICC) \citep{kish1965,donner1981}. Although this DE is commonly used in sample size calculations for cluster RCTs \citep{campbell2014, rutterford2015}, it was derived for comparisons of means. We will show that the sample size for an adequately powered individual RCT based on Whitehead's formula \citep{whitehead1993} can be inflated for cluster data using a DE that incorporates the rank ICC \citep{tu2023} -- a rank-based correlation measuring the degree of similarity within clusters. Therefore, we provide a unifying sample size calculation framework for the design of RCTs with response variables that are continuous, ordinal, or a mixture of the two (e.g., data with detection limits) and for both individual- and cluster-randomized trials. 

This paper is structured as follows. In Section 2, we briefly review some rank-based tests and models for individual and cluster RCTs. In Section 3, we first extend Whitehead's sample size calculations \citep{whitehead1993} for ordinal outcomes in individual RCTs to continuous outcomes. We then introduce a DE that incorporates the rank ICC as an inflation factor and propose new sample size calculations for cluster RCTs with skewed or ordinal outcomes using this inflation factor. In Section 4, we conduct simulations to evaluate the performance of our sample size calculations. In Section 5, we illustrate the use of our approach in the design of three RCTs: an individual RCT with a skewed continuous outcome, a cluster RCT with a skewed continuous outcome, and a non-inferiority cluster RCT with an irregularly distributed count outcome. Section 6 provides a discussion. Additional information is available in the Supplementary Materials.

\section{Review of rank-based methods for individual and cluster RCTs}
\label{s:review}
Rank-based methods are usually used to analyze skewed or ordinal data, given their nonparametric nature and robustness to the shape of the distribution. The Wilcoxon rank-sum test, equivalent to the Mann–Whitney U test, is a widely used rank-based approach for evaluating treatment effects with skewed or ordinal data in the absence of clustering \citep{mann1947, lehmann1975}. The parameter of the Wilcoxon rank-sum test can be formulated in terms of the probabilistic index, $\theta = P(X < Y) + P(X = Y)/2$, where $X$ and $Y$ are random variables from two different groups \citep{hollander1999}. With continuous $X$ and $Y$, $\theta = P(X < Y)$. The hypotheses formulated in terms of $\theta$ are $H_0: \theta =  1/2$ vs. $H_1: \theta \neq 1/2$. The estimator of $\theta$ from an individually randomized (IR) controlled trial, denoted as $\hat{\theta}_{\text{\scriptsize IR}}$, is equal to the Mann-Whitney U statistic divided by the product of the sample sizes of the two arms, $n_{\text{\scriptsize IR}_0}$ and $n_{\text{\scriptsize IR}_1}$. Specifically, given independent data ($\{x_i: i=1,...,n_{\text{\scriptsize IR0}}\}$ and $\{y_s: s=1,...,n_{\text{\scriptsize IR1}}\}$) from the two arms of an individual RCT, $\theta$ is $\hat{\theta}_{\text{\scriptsize IR}} = \sum\limits_{i}\sum\limits_{s} U(x_i, y_s)/(n_{\text{\scriptsize IR0}}n_{\text{\scriptsize IR1}})$, where $U(a,b)=1$ if $a<b$; $1/2$ if $a=b$, and $0$ if $a>b$. 

Proportional odds (PO) models are commonly applied in the analysis of ordinal outcomes \citep{mccullagh1980}. PO models can also be fit as a robust and rank-based analysis of continuous outcomes, where each unique continuous outcome is treated as an ordinal category \citep{liu2017}. Specifically, let $Z$ be the outcome ($X$ or $Y$) and $G$ be the treatment assignment indicator (e.g., 0 if $Z$ is $X$ and 1 if $Z$ is $Y$). Then the PO model is $\text{logit}\{P(Z \leq z | G)\} = \eta(z) + \delta G$, where $\delta$ is the log odds ratio (OR) for the treatment effect and the intercept function $\eta(z)$ is estimated using a step function resulting in a parameter vector of length one fewer than the number of unique continuous outcomes. Since only the treatment indicator is in the model, we refer to this model as the \textit{unadjusted PO model}. With continuous data, considering all possible ordered dichotomizations of $Z$, the OR is interpreted as the relative odds based on treatment group $G$ of being in a higher category of any dichotomized $Z$. It has been shown that fitting a PO model to continuous data is equivalent to fitting a semiparametric linear transformation model \citep{zeng2007,liu2017}. Whitehead \citep{whitehead1993} has shown that the test statistic for treatment effect, based on the efficient score statistic derived from an unadjusted PO model, is exactly equal to a version of the Mann-Whitney U test statistic presented by Siegel \citep{siegel1956}. That is, unadjusted PO models with a single binary covariate are essentially Wilcoxon rank-sum/Mann-Whitney U tests. Additionally, there is a numerical relationship between $\theta$ and the log OR regarding the treatment effect in unadjusted PO models \citep{de_neve2019}: $\theta =  \exp(\delta)[\exp(\delta)-\delta-1]/(\exp(\delta)-1)^2$, where $\delta$ denotes the log OR.

To account for clustering, Rosner et al. \citep{rosner2003} developed the clustered Wilcoxon rank-sum test, which incorporates a correction to the variance of the Wilcoxon rank-sum test statistic. The clustered Wilcoxon rank-sum test can also be expressed in terms of $\theta$. The definition of $\theta$ for clustered data is the same as that for unclustered data: $\theta = P(X < Y) + P(X = Y)/2$, where $X$ and $Y$ are now random variables from hierarchical distributions. This $\theta$ has been used in power and sample size estimation for the clustered Wilcoxon rank-sum test with continuous data \citep{rosner2011}. Given clustered data ($\{x_{ij}:i=1,...,m_{0}, j=1,...,k_{0i}\}$ and $\{y_{st}:s=1,...,m_{1}, t=1,...,k_{1s}\}$) with the cluster size $k$ ranging between $k_{\text{min}}$ and $k_{\text{max}}$, an estimator of $\theta$ is 
$\hat{\theta} = \sum\limits_{k=k_{\text{min}}}^{k_{\text{max}}} w^{(k)} \hat{\theta}^{(k)} / \sum\limits_{k=k_{\text{min}}}^{k_{\text{max}}} w^{(k)}$, where $w^{(k)} = 1 / \text{var}(\hat{\theta}^{(k)})$, $\hat{\theta}^{(k)} = \sum\limits_{i}\sum\limits_{j}\sum\limits_{s}\sum\limits_{t} U(x_{ij}^{(k)}, y_{st}^{(k)})/(n_0^{(k)} n_1^{(k)})$, $x_{ij}^{(k)}$ and $y_{st}^{(k)}$ are from clusters of size $k$, and $n_0^{(k)}$ and $n_1^{(k)}$ are the sample sizes of clusters of size $k$ \citep{rosner2006}. Because the clustered Wilcoxon rank-sum test is a large-sample approach that groups clusters by cluster sizes, this test could have challenges when there are not many clusters of the same size. This limitation is seen in our simulations (see Section \ref{s:simulations}). 

PO models have also been extended to handle clustered ordinal or continuous outcomes \citep{heagerty1996, parsons2006, tian2023}, employing GEE-based estimation. Commonly used working correlation structures include independent, exchangeable, and first-order autoregressive (AR1) correlations. Tian et al. \citep{tian2023} demonstrated that clustered continuous outcomes could also be analyzed using PO GEE-based methods. Fitting the unadjusted PO model to the continuous outcome and then fixing the standard error of model coefficients using a Huber-White sandwich estimator of the covariance to correct for within-cluster correlation is straightforward to implement and is equivalent to fitting a GEE with an independent working correlation structure. Exchangeable/AR1 working correlation structures can be statistically more efficient than independent working correlation in some settings with continuous outcomes but are more computationally burdensome. In this paper, we focus on PO models estimated with GEE with independent working correlation, which we henceforth refer to as \textit{cluster PO models} for simplicity. 

\section{Sample size calculations}
\label{s:calculation}

Since PO models can be used to analyze ordinal and continuous outcome data, both independent and clustered, sample size calculations based on PO models can be used to design individual and cluster RCTs for both ordinal and continuous outcomes. Table \ref{tab:summary} summarizes analysis methods and sample size formulas for ordinal versus continuous, and independent versus clustered outcomes. We present the details of our approach below. 

\begin{table*}[t]
    \centering
    \caption{Summary of analysis methods and sample size calculations for ordinal and continuous outcome data that are independent versus clustered.}
    \tabcolsep=0pt%
    \begin{tabular*}{\textwidth}{p{4cm}p{6cm}p{5.5cm}}
        \hline 
        & Ordinal & Continuous \\
        \hline
       Independent data  & &  \\
       \:\:\:\:\:\: Analysis & PO models / Wilcoxon tests & PO models / Wilcoxon tests\\
       \:\:\:\:\:\: Design & Formula (\ref{nsrs}) & Formula (\ref{nsrs_cont}) \\
        \hline
       Clustered data  & &  \\
       \:\:\:\:\:\: Analysis & Cluster PO models / clustered \:\:\:\: Wilcoxon tests & Cluster PO models / clustered Wilcoxon tests \\
       \:\:\:\:\:\: Design & Formula (\ref{noc}) & Formula (\ref{ncont}) \\
       \hline
    \end{tabular*}
    \label{tab:summary}
\end{table*}

\subsection{Individual RCTs}
\label{s:calculation_individual}
For simplicity, we refer to the two arms in RCTs as the control and experiment arms. Under individual randomization (IR), Whitehead \citep{whitehead1993} provided a sample size calculation formula for ordinal outcomes using an approximation of the variance of the score statistic derived from the likelihood of the observed ranks, referred to as the ``marginal likelihood'', of an unadjusted PO model. Let $n_{\text{\scriptsize IR}}$ denote the total sample size for an individual RCT and $A > 0$ denote the allocation ratio of the control arm to the experiment arm. For a two-sided significance level at $\alpha$ and power at $1-\beta$, Whitehead's formula is 
\begin{equation}
n_{\text{\scriptsize IR}} = \frac{3(A+1)^2(Z_{1-\alpha/2}+Z_{1-\beta})^2/\delta^2}{A(1 - \sum_{l=1}^L \bar{\pi}_{l}^3)}, \label{nsrs}
\end{equation}
where $\bar{\pi}_l$ is the mean proportion expected in the $l$th ordinal category and calculated as $\bar{\pi}_l = (\pi_{0l}+\pi_{1l})/2$, $\pi_{0l}$ and $\pi_{1l}$ are the proportions for the control and experiment groups, $L$ is the total number of ordered categories, and $\delta$ denotes the log OR of experiment versus control in the unadjusted PO model. With (\ref{nsrs}), the sample sizes for the experiment and control arms can be calculated as $\ceil*{ n_{\text{\scriptsize IR}}/(A+1)}$ and $\ceil*{An_{\text{\scriptsize IR}}/(A+1)}$, respectively, where $\ceil*{\cdot}$ denotes rounding up. For one-sided tests, we use $Z_{1-\alpha}$ instead of $Z_{1-\alpha/2}$. 

Continuous outcomes are also ordinal, and as continuous data can be analyzed as if it were ordinal \citep{liu2017}, the formula (\ref{nsrs}) for ordinal outcomes can also be applied to continuous outcomes. Since truly continuous outcomes have no ties, the proportion for each ordinal category is $\bar{\pi}_{l} = 1/n_{\text{\scriptsize IR}}$. The score statistic underlying Whitehead's formula is the same as the score statistic derived from a CPM of the continuous outcome. Whitehead's variance approximation for the score statistic continues to hold for continuous outcomes if $\bar{\pi}_{l}$ is replaced with $1/n_{\text{\scriptsize IR}}$, and in fact, the approximation is better for continuous data than it is for ordinal data with only a few categories (see Supplementary Materials Appendix A for details). This provides an analytical rationale for the extension of Whitehead's formula to continuous data. Specifically, we replace $\bar{\pi}_{l}$ in (\ref{nsrs}) with $1/n_{\text{\scriptsize IR}}$, resulting in
$$ n_{\text{\scriptsize IR}} = \frac{3(A+1)^2(Z_{1-\alpha/2}+Z_{1-\beta})^2/\delta^2}{A(1 - \sum_{l=1}^{\textstyle{n}_{\text{\scriptsize IR}}} 1/n_{\text{\scriptsize IR}}^3)},$$ and solve for positive $n_{\text{\scriptsize IR}}$. The solution, which is the sample size for individual RCTs with continuous outcomes, is then 
\begin{equation}
 n_{\text{\scriptsize IR}} = \sqrt{1+S^2}+S, \label{nsrs_cont}   
\end{equation}
where $S=3(A+1)^2(Z_{1-\alpha/2}+Z_{1-\beta})^2/(2A\delta^2)$.
The effect size in (\ref{nsrs_cont}) is specified in terms of the log OR, $\delta$, where the OR for continuous outcomes is the relative odds of being in a higher category of any dichotomization of the outcome (see Section \ref{s:review}). The effect size can also be specified in terms of the probabilistic index, $\theta =  \exp(\delta)[\exp(\delta)-\delta-1]/(\exp(\delta)-1)^2$. The calculation provided in (\ref{nsrs_cont}) is rank-based, and therefore it is robust to skewness, extreme values, and any data transformations. 

\begin{figure*}
     \centering
     \subfloat[Sample size per arm]{\includegraphics[width=0.5\textwidth]{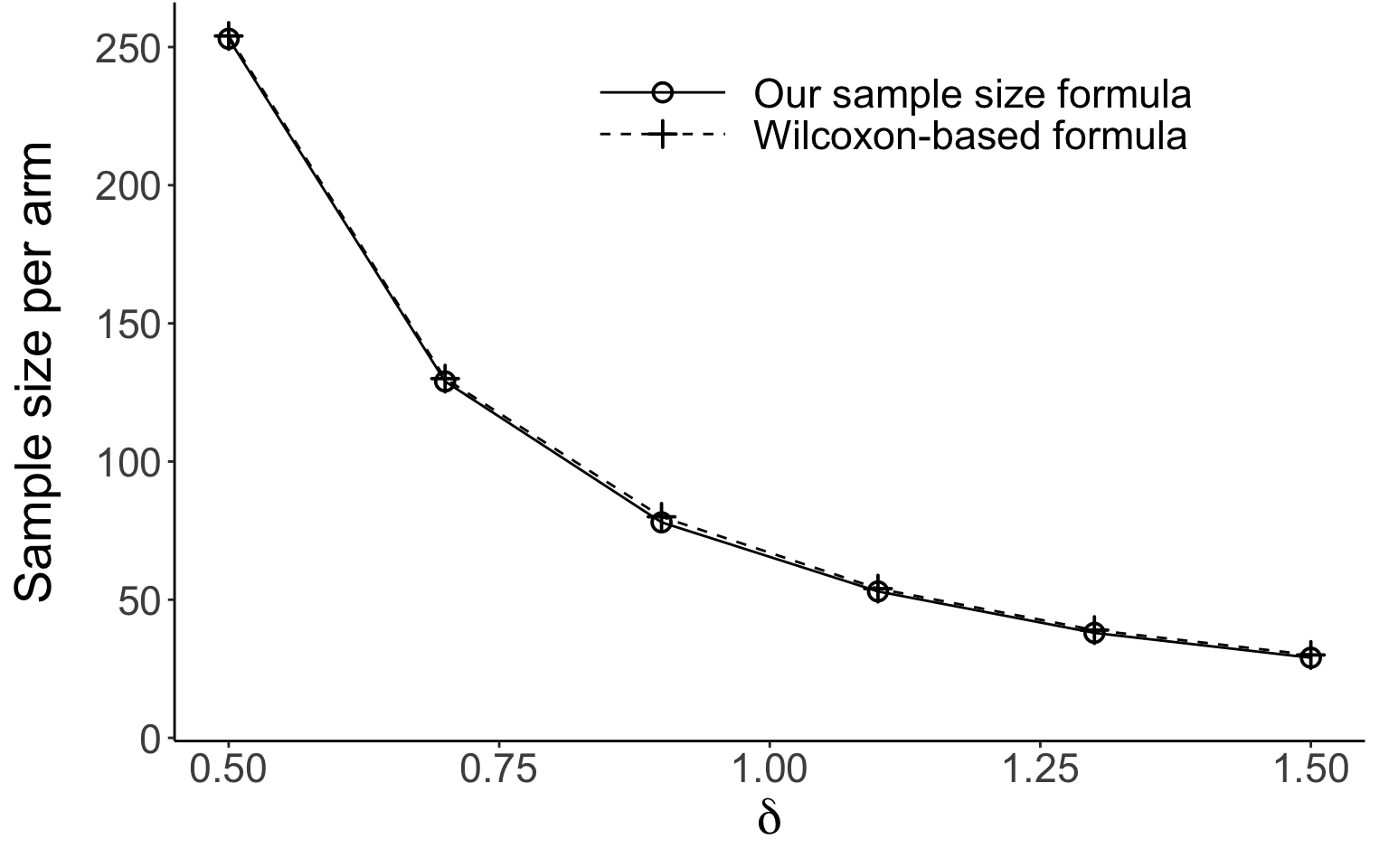}}
     \quad
    \subfloat[Power]{\includegraphics[width=0.5\textwidth]{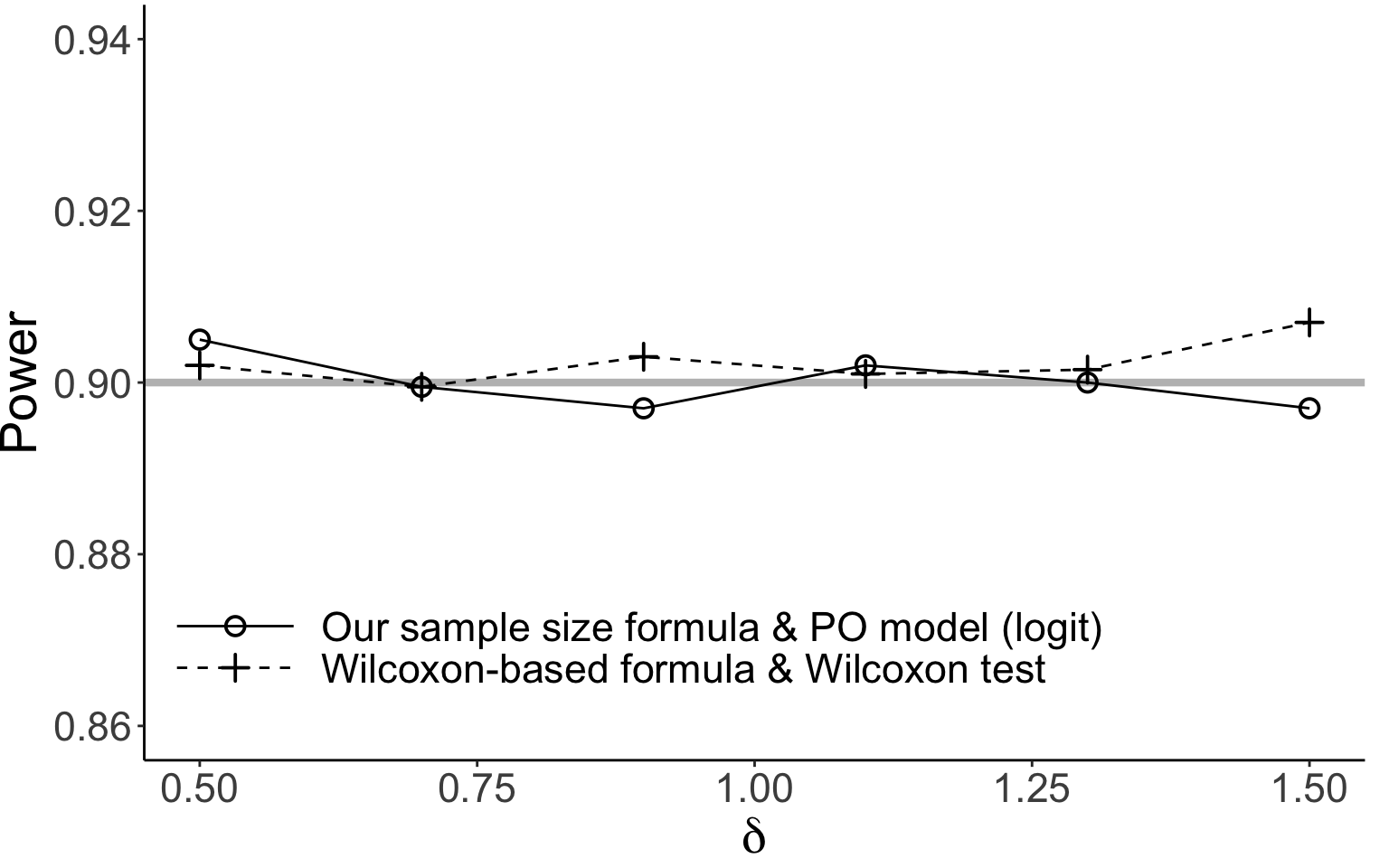}}
     \caption[]{\centering Sample size per arm to achieve 90\% power and observed power for independent continuous data over different treatment effects.}
     \label{fig:size_power_ind}
\end{figure*}

There is an alternative rank-based sample size calculation approach, derived from Wilcoxon rank sum tests and used for continuous outcomes \citep{rosner2009}. However, this sample size calculation approach is complex and lacks a closed form. We show via simulation that the sample sizes obtained by (\ref{nsrs_cont}) are very close to those obtained by the approach based on Wilcoxon rank-sum tests (Section 4, Figure \ref{fig:size_power_ind}).
 
\subsection{Cluster RCTs}
\label{s:cluster}
\subsubsection{Design effect of cluster RCTs}
\label{s:de}
The design effect (DE) is often used as an inflation factor for sample size calculations in cluster RCTs. It was initially introduced by Kish \citep{kish1965} as a measure of the expected impact of a sampling design on the variance of an estimator. Subsequently, it was applied by Donner et al. \citep{donner1981} to inflate sample sizes calculated under individual randomization to achieve the required statistical power under cluster randomization. The DE of cluster RCTs with respect to an estimator $T$ is defined as
$$D_{\text{eff}}(T) = \frac{\text{var}(T)}{\text{var}(T_{\text{\scriptsize IR}})},$$
where $\text{var}(T)$ is calculated under cluster randomization and $\text{var}(T_{\text{\scriptsize IR}})$ is calculated under individual randomization with the same number of observations.

Let $\rho_I$ denote the intraclass correlation coefficient (ICC) and $\rho_I = corr(X_{ij},X_{ij'})$, where $(X_{ij},X_{ij'})$ is a random pair from a random cluster \citep{fisher1925}. The DE of cluster RCTs with a common cluster size $k$ for the mean $\bar X$ is \citep{kish1965,kish1987}
\begin{equation}
     D_{\text{eff}}(\bar X) = 1+\rho_I(k-1). \label{derho}
\end{equation}
The DE in (\ref{derho}) is often used as the inflation factor for sample size calculations in cluster RCTs \citep{campbell2014, rutterford2015}. Let $n_{\text{\scriptsize IR}}$ denote the total sample size for an individual RCT with the targeted power. Then the total sample size for a cluster RCT with the same power is calculated as
$$n = n_{\text{\scriptsize IR}}D_{\text{eff}}(\bar X) = n_{\text{\scriptsize IR}}\{1+\rho_I(k-1)\}.$$
However, estimation of $\rho_I$ is sensitive to extreme values and skewed distributions, and it depends on the scale of the data. $\rho_I$ also lacks a clear definition when applied to ordinal data. While ordinal regression models with random effects may be used to estimate variance components, the total variance remains undefined unless numbers are assigned to levels of the ordinal response \citep{denham2016}. Hence, the DE in (\ref{derho}) may not be applicable to skewed or ordinal data. 

The DE for the clustered Wilcoxon rank-sum statistic, $\hat{\theta}$, is $D_{\text{eff}}(\hat{\theta}) = \text{var}(\hat{\theta})/\text{var}(\hat{\theta}_{\text{\scriptsize IR}})$. Although closed-form formulas for $\text{var}(\hat{\theta})$ and $\text{var}(\hat{\theta}_{\text{\scriptsize IR}})$ can be derived under the null hypothesis ($\theta=1/2$), their derivation under the alternative hypothesis, which is relevant to sample size calculations, requires additional assumptions \citep{lehmann1975, rosner2011}. Specifically, the closed-form formulas for $\text{var}(\hat{\theta})$ and $\text{var}(\hat \theta_{\text{\scriptsize IR}})$ require continuous distributions, and for $\text{var}(\hat{\theta})$, involves specifying the ICC after an implicit transformation such that the transformed data are normally distributed.
Hence, $D_{\text{eff}}(\hat{\theta})$ using the variances of independent and clustered Wilcoxon rank-sum statistics does not result in a simple inflation factor in sample size calculations for skewed continuous outcomes, and it is also not applicable to ordinal outcomes. 

We consider an alternative inflation factor that closely approximates $D_{\text{eff}}(\hat{\theta})$ in most scenarios and is also applicable to ordinal outcomes. Let $F$ be a CDF, $F(x-)=\lim_{t \uparrow x} F(t)$, and $F^*(x) = \{F(x) + F(x-)\}/2$. If the distribution is continuous, then $F^*(x) = F(x)$. If the distribution is discrete or mixed, $F^*(x)$ corresponds to the population versions of ridits or midranks \citep{Bross1958}. Let $\gamma_I$ denote the rank ICC, which is a rank-based correlation measuring the degree of within-cluster similarity. It is defined as follows,
\begin{align*}
    \gamma_I = corr\{F^*(X_{ij}), F^*(X_{ij'})\},
\end{align*} 
where $(X_{ij}, X_{ij'})$ is a random pair from a random cluster \citep{tu2023}. $\gamma_I$ is insensitive to extreme values and skewed distributions, and it does not depend on the scale of data. It is also applicable to ordinal data and is easily computed. We analytically show that when $\theta=1/2$ and the numbers of clusters in the two arms greatly exceed cluster sizes,
\begin{equation}
    D_{\text{eff}}(\hat{\theta}) \approx 1 + \gamma_{I}(k-1). \label{de_approx}
\end{equation}
See Supplementary Materials Web Appendix B for details. We also show via simulations that in other scenarios, (\ref{de_approx}) remains a good approximation, except for very large $\gamma_I$ (Web Figure 1). Given the nonparametric nature of $\gamma_I$ and the approximation between $1 + \gamma_I(k-1)$ and $D_{\text{eff}}(\hat{\theta})$, $1 + \gamma_I(k-1)$ can serve as an inflation factor in sample size calculations for skewed or ordinal outcomes. This inflation factor is as simple as the conventional $1+\rho_I(k-1)$ but it is more robust in the context of skewed data and is applicable to ordinal data. 

\subsubsection{Sample size calculations for cluster RCTs}
\label{s:calculation_cluster}
We extend Whitehead's sample size calculation for individual RCTs to cluster RCTs by using $1+\gamma_I(k-1)$ to inflate $n_{\text{\scriptsize IR}}$ in (\ref{nsrs}). With ordinal outcomes, the total sample size of a cluster RCT with a two-sided significance level $\alpha$ and power $1-\beta$ is calculated as 
\begin{align*}
n & = n_{\text{\scriptsize IR}}\{1+\gamma_I(k-1)\} \\
&  = \frac{3(A+1)^2(Z_{1-\alpha/2}+Z_{1-\beta})^2/\delta^2}{A(1 - \sum_{l=1}^L\bar{\pi}_l^3)}\{1+\gamma_I(k-1)\}. \numberthis \label{noc} 
\end{align*}
With (\ref{noc}), the sample sizes for the experiment and control arms can be easily calculated as $n_1 = n/(A+1)$ and $n_0 = An/(A+1)$, respectively. The numbers of clusters for the two arms to ensure at least $1-\beta$ power are then calculated as $\ceil*{n_1/k}$ and $\ceil*{n_0/k}$, where $\ceil*{\cdot}$ denotes rounding up. 

Similar to the extension from (\ref{nsrs}) to (\ref{nsrs_cont}), we can also apply the sample size formula (\ref{noc}) to continuous outcomes in cluster RCTs. For continuous outcomes, the proportion at each ordinal category is $\bar{\pi}_{l}= 1/n$. We plug this proportion into (\ref{noc}), 
\begin{equation*}
    n = \frac{3(A+1)^2(Z_{1-\alpha/2}+Z_{1-\beta})^2/\delta^2}{A(1 - \sum_{l=1}^{n} 1/n^3)}\{1+\gamma_I(k-1)\}, 
\end{equation*}
and solve for positive $n$. The solution, which is the total sample size for cluster RCTs with continuous outcomes, is then
\begin{equation}
n = \sqrt{1+S^2 \{1+\gamma_I(k-1)\}^2} + S\{1+\gamma_I(k-1)\},\label{ncont} 
\end{equation}
where, as noted previously, $S=3(A+1)^2(Z_{1-\alpha/2}+Z_{1-\beta})^2/(2A\delta^2)$, $\delta$ denotes the log OR in the unadjusted cluster PO model, and $A$ is the allocation ratio. When $\gamma_I$ is 0 or the cluster size $k$ is 1, (\ref{noc}) and (\ref{ncont}) simplify to (\ref{nsrs}) and (\ref{nsrs_cont}), respectively, which are for individual RCTs. 

In practice, there are situations where the number of clusters is predetermined and the goal is to calculate the cluster size. In such cases, our method can also calculate cluster sizes with an appropriately predetermined number of clusters. Let $m$ denote the total number of clusters. With ordinal outcomes, the calculation for the cluster sizes is derived from formula (\ref{noc}), 
\begin{equation}
    k = \ceil*{\frac{2S(1-\gamma_I)}{m(1 - \sum_{l=1}^L\bar{\pi}_L^3) - 2\gamma_IS}},
\end{equation}
when $m(1 - \sum_{l=1}^L\bar{\pi}_L^3) > 2\gamma_IS$. With continuous outcomes, the calculation for the cluster sizes is derived from formula (\ref{ncont}),
\begin{equation}
    k =\ceil*{ \sqrt{\frac{1}{m(m - 2\gamma_IS)} + \frac{S^2(1-\gamma_I)^2}{(m-2\gamma_IS)^2}}  + \frac{S(1-\gamma_I)}{m - 2\gamma_IS}},
\end{equation}
when $m > 2\gamma_IS$. When $m(1 - \sum_{l=1}^L\bar{\pi}_L^3) \leq 2\gamma_IS$ for ordinal outcomes or $m \leq 2\gamma_IS$ for continuous outcomes, there is no finite cluster size to achieve the desired power. 

Typically, $1+\rho_I(k-1)$ is used to inflate the sample size for an adequately powered individual RCT based on two-sample t-tests. It is expected that, under the assumption of normality, the sample sizes calculated from this conventional approach would be smaller than those from our calculations. However, we show that under normality, if the allocation ratio is $1$ or the outcome variances of both arms are equal, the sample sizes calculated from this conventional approach are very close to those from our calculations. Details on the derivations are available in the Supplementary Materials (Web Appendix C). This implies that there is little penalty in terms of additional sample size for not assuming normality and instead using our more robust sample size calculations. 

\section{Simulations}
\label{s:simulations}
\subsection{Individual RCTs}
We performed simulations to compare our sample size calculations (formula (\ref{nsrs_cont})) with the sample size calculations based on the Wilcoxon rank-sum test for continuous outcomes. These two approaches have been developed from unadjusted PO models and the Wilcoxon rank-sum test, respectively. The power of each calculation approach was therefore estimated with the test from which it was developed. Let $X_i$ and $Y_j$ denote observations in the control and experiment groups, respectively. $\{X_{i}\}$ are independent and identically distributed (i.i.d.) following a logistic distribution with a location parameter of 0 and a scale parameter of 1 and $\{Y_{i}\}$ are i.i.d. following a logistic distribution with a location parameter of $\delta$ and a scale parameter of 1, where $\delta$ is the log OR of the treatment effect in the unadjusted PO model and $\delta$ varies between $0.5$ and $1.5$. The two-sided significance level was set to 0.05 and the desired power was set to 0.9. Power simulations were conducted 1,000 times for each value of $\delta$. The simulation results are summarized in Figure \ref{fig:size_power_ind}. The two calculation approaches yielded nearly identical sample sizes, with both achieving powers approximately equal to the target value of 0.9. 

\subsection{Cluster RCTs}
\subsubsection{Continuous Outcomes}
We generated cluster RCT data using two additive models: $X_{0ij} = U_{Xi}+R_{Xij}$ and $Y_{0ts}=U_{Yt}+R_{Yts}$, where $U_{Xi} \stackrel{i.i.d}{\sim} N(0,\rho_{I})$, $R_{Xij} \stackrel{i.i.d}{\sim} N(0,1-\rho_{I})$, $U_{Yt} \stackrel{i.i.d}{\sim} N(\mu,\rho_{I})$, $R_{Yts} \stackrel{i.i.d}{\sim} N(0,1-\rho_{I})$, and $\rho_{I}$, the ICC on the latent scale, varies over $[0,0.9]$. The corresponding rank ICC is $\gamma_I = 6\arcsin(\rho_{I}/2)/\pi$ \citep{Pearson1907}. Let $X_{ij}$ and $Y_{ts}$ denote observations in the control and experiment groups, respectively, $X_{ij}=\exp(X_{0ij})$ and $Y_{ts}=\exp(Y_{0ts})$. We considered different magnitudes of the treatment effect: $\delta=\{0, 0.1, 0.5,1,1.5\}$, where $\delta$ is the log OR of the treatment effect in the unadjusted PO model. As described in Section 2, the value of $\theta$ can be calculated from $\delta$ \citep{de_neve2019}. We then can compute the value of $\mu$ from $\theta$ by $\mu = \sqrt{2}\Phi^{-1}(\theta)$, because $\theta = P(X-Y<0) = P((V+\mu)/\sqrt{2} < \mu/\sqrt{2}) = \Phi(\mu/\sqrt{2})$ where $V \sim N(-\mu,2)$. In addition, we explored two design scenarios with predetermined equal cluster sizes: one with a small cluster size of 5, and another with a large cluster size of 50. The allocation ratio of the control arm to the treatment arm was set to 1. The two-sided significance level was set to 0.05 and the desired power was set to 0.9. All power simulations were conducted 1,000 times for each setting. 

\begin{figure*}[!t]
     \centering
     \subfloat[Cluster sizes = 5]{\includegraphics[width=0.7\textwidth]{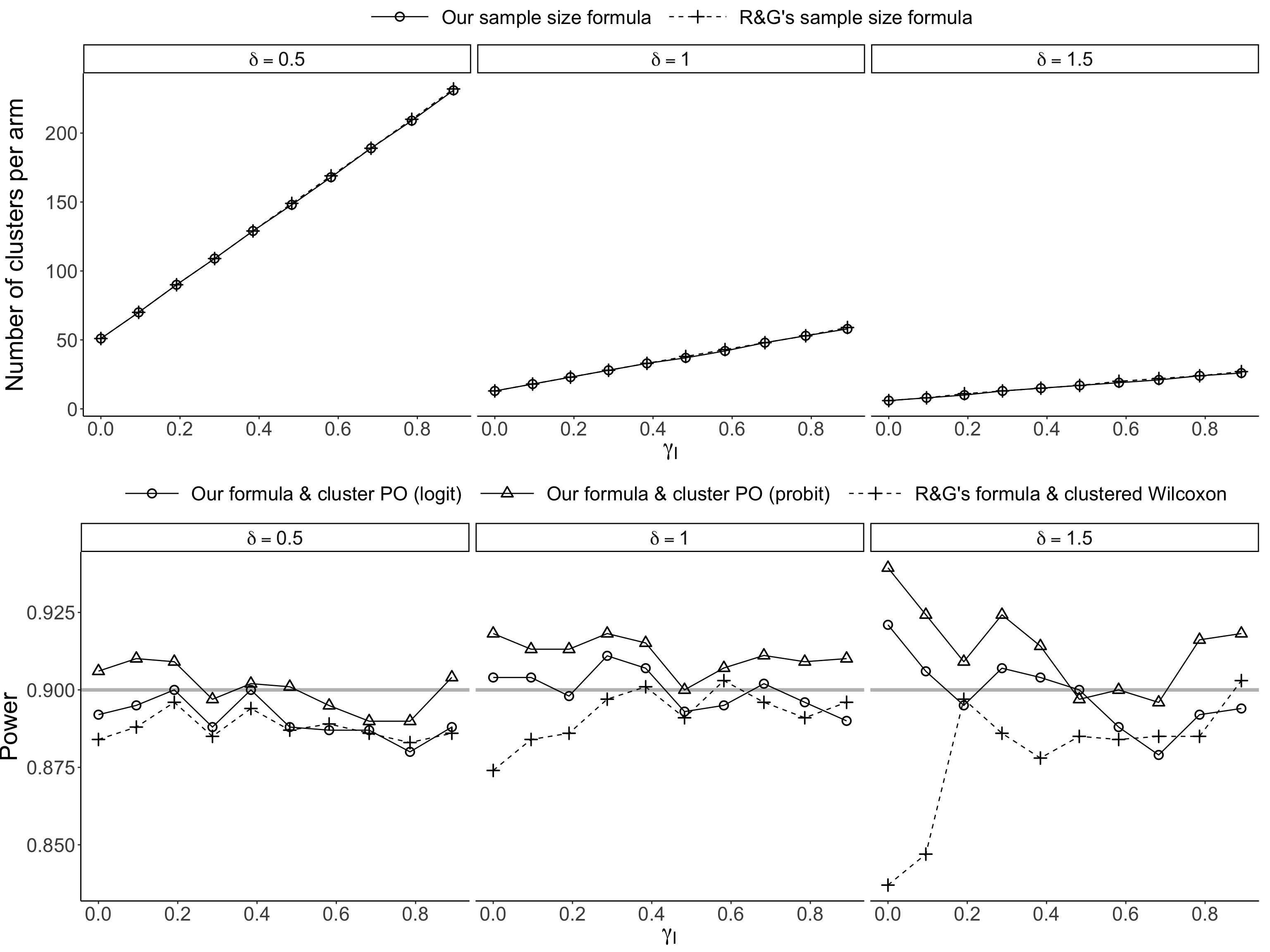} \label{fig:size_powerk5}}
     \quad
    \subfloat[Cluster sizes = 50]{\includegraphics[width=0.7\textwidth]{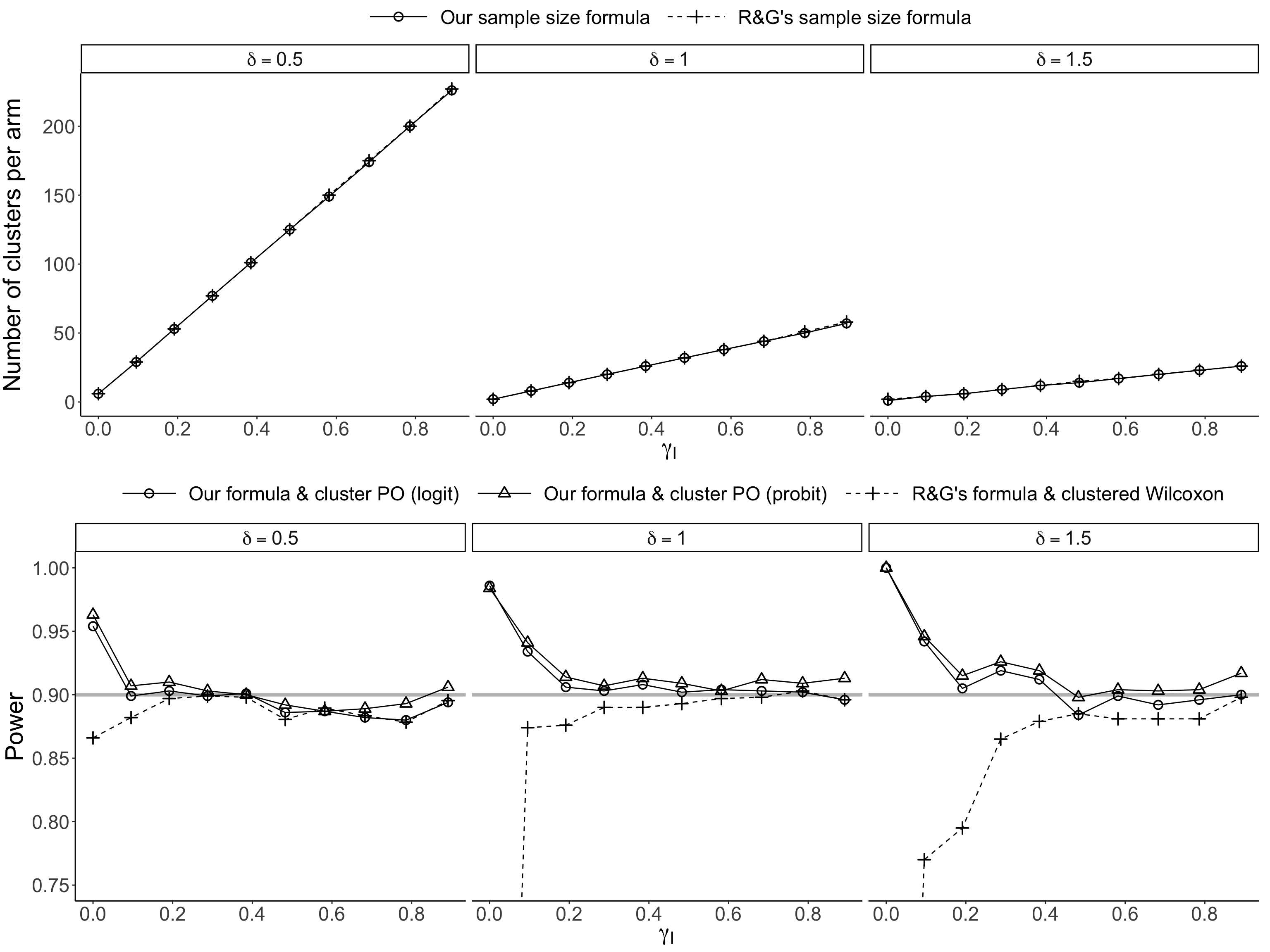}  \label{fig:size_powerk50}}
     \caption[]{Number of clusters per arm and observed power for continuous data with predetermined cluster sizes of 5 and 50 with the number of clusters selected as that needed to achieve 90\% power. ``R\&G's formula'' represents Rosner and Glynn's sample size formula based on clustered Wilcoxon tests. When cluster sizes $=50$, $\gamma_I=0$, and $\delta=1$ or $1.5$, the estimated power of the ``R\&G's formula \& clustered Wilcoxon'' was 0.}
     \label{fig:size_power_cnt}
\end{figure*}

We first evaluated the power of our sample size calculations using cluster PO models with the correct probit link function (because the data were generated with normally distributed latent variables) and a misspecified logit link function. As shown in Figure \ref{fig:size_power_cnt}, the power under both link functions was approximately equal to the target of 0.9, with the power under the probit link being slightly greater. When $\delta$ was small, the power under the logit link was slightly below 0.9, perhaps due to link function misspecification.
 
Figure \ref{fig:size_power_cnt} also compares our sample size calculations with Rosner and Glynn's sample size calculations with respect to power. Similar to the two compared calculation approaches for individual RCTs, these two approaches for cluster RCTs have also been developed differently, stemming from unadjusted PO models and clustered Wilcoxon rank-sum tests, respectively. The power of each calculation approach was therefore estimated with the test from which it was developed: unadjusted cluster PO models with logit link for our calculations, and clustered Wilcoxon rank-sum tests for Rosner and Glynn's calculations.  

In summary, our sample size calculations led to simulation results with the power approximately equal to the target 0.9 in most scenarios we considered. The sample sizes obtained by the two calculation approaches were very close and both increased approximately linearly with $\gamma_I$ (Figure \ref{fig:size_power_cnt}). This suggests that, both calculation approaches have an approximately linear relationship with $\gamma_I$, even though Rosner and Glynn's calculations are more complicated. When the cluster size was large ($k=50$) and $\gamma_I$ was close to 0, the power of our sample size calculations was greater than 0.9 (Figure \ref{fig:size_powerk50}). This is because this predetermined cluster size exceeded the required number of individuals (e.g., when $\gamma_I=0$, fewer than 50 subjects in each arm are needed for 90\% power and $\delta=1.5$). Rosner and Glynn's calculations had poor power in situations with small $\gamma_I$ and large cluster sizes. This may be due to the poor performance of the clustered Wilcoxon rank-sum test with small numbers of clusters, as this test was proposed as a large-sample approach \citep{rosner2003}.

Furthermore, the power of our sample size calculations was slightly higher than that of Rosner and Glynn's calculations in most scenarios. This difference might be attributed to differences between unadjusted cluster PO models and clustered Wilcoxon rank-sum tests. Thus, we performed a comparison between unadjusted cluster PO models and clustered Wilcoxon rank-sum tests under the null with $\delta=0$ ($\theta=1/2$). The cluster size was fixed at $k=5$ and the number of clusters varied as 20, 50, 100, and 300. Simulation results are summarized in Web Figure 2. Although the type I error rate was close to the nominal 0.05 level for both procedures at all sample sizes, unadjusted PO models had a slightly higher type I error rate than clustered Wilcoxon rank-sum tests. This difference tended to diminish as $\gamma_I$ increased or the number of clusters increased. This difference in the type I error rate might explain why our sample size calculations had a slightly higher power in the simulations mentioned above, especially with smaller numbers of clusters.

\begin{figure*}
     \centering
    \includegraphics[width=0.85 \textwidth]{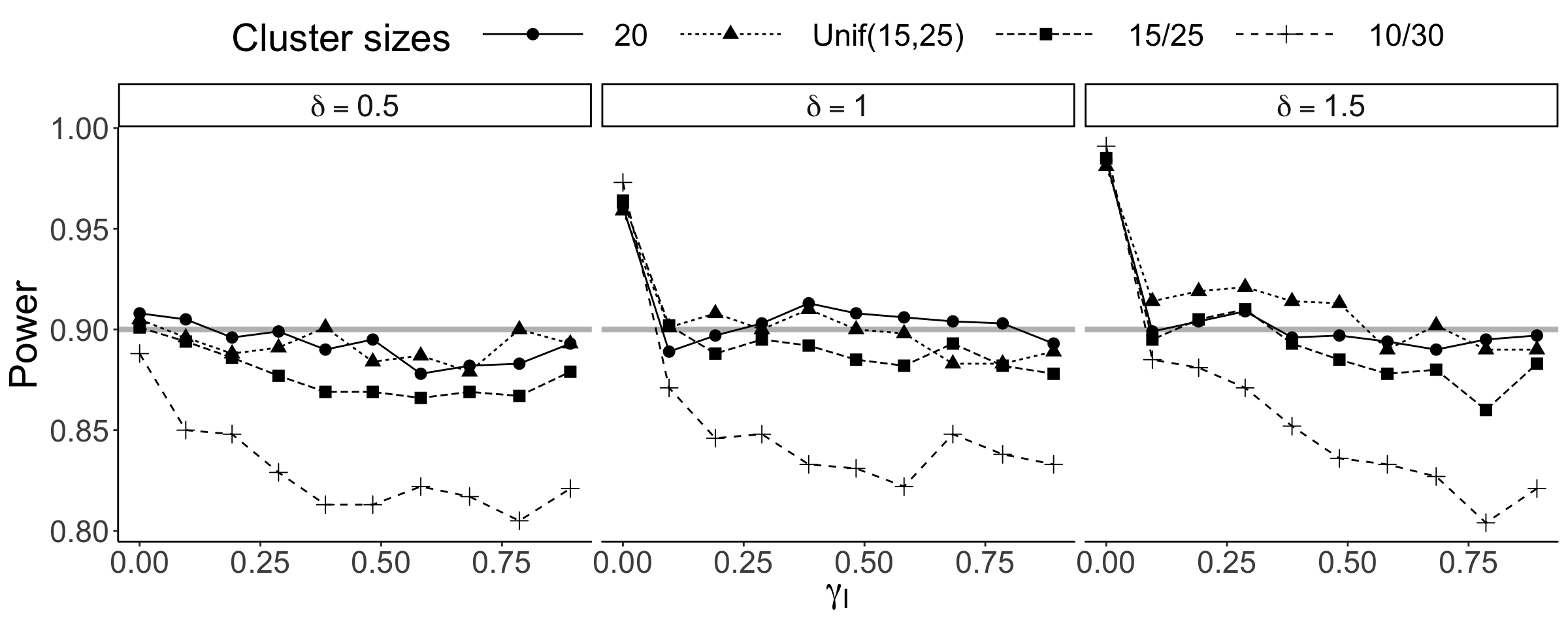} 
     \caption[]{Number of clusters per arm calculated based on predetermined equal cluster sizes of 20, and observed power under equal or unequal cluster sizes. The number of clusters per arm was calculated as that needed to achieve 90\% power as a function of the log odds ratio, $\delta$, and the rank ICC, $\gamma_I$, based on predetermined equal cluster sizes of 20. ``Unif(15,25)'', ``15/25'', and ``10/30'' represent cluster sizes in actual sample data uniformly ranging from 15 to 25, half each of 15 and 25, and half each of 10 and 30, respectively.}
     \label{fig:unequalk}
\end{figure*}

In practice, when designing a cluster RCT, it is common to predetermine equal cluster sizes to calculate sample sizes, but the cluster sizes after accrual may be unequal. Therefore, we conducted simulations to evaluate the performance of our sample size calculations in such cases. The data generation process involved first computing the numbers of clusters with predetermined equal cluster sizes, and then generating data with the computed numbers of clusters and sample sizes but unequal cluster sizes. The predetermined cluster size for sample size calculations was set to 20. We explored various configurations of cluster sizes of actual sample data: (a) equal at 20; (b) uniformly ranging from 15 to 25; (c) half each of 15 and 25; (d) half each of 10 and 30. The power was obtained via simulations based on cluster PO models. The results are summarized in Figure \ref{fig:unequalk}. The powers of (a) and (b) were very close, while the power of (c) was slightly smaller. The power of (d) was much smaller than the others. These simulations suggest that if the unequal cluster sizes in actual sample data do not differ much from the predetermined equal cluster sizes, our sample size calculations remain robust, but if the difference is very large, our trial might be underpowered. When clustered Wilcoxon rank-sum tests were fit to the same data, power was also low with extreme cluster size imbalance (Web Figure 3). Interestingly, the clustered Wilcoxon rank-sum test had especially low power when cluster sizes were uniformly distributed between 15 to 25; this test appears to have challenges when there are few clusters of the same cluster size because the algorithm performs computations within equal-sized clusters. 

\subsubsection{Ordinal Outcomes}
\begin{figure*}[!t]
    \centering
     \subfloat[cluster sizes = 5]{\includegraphics[width=0.8\textwidth]{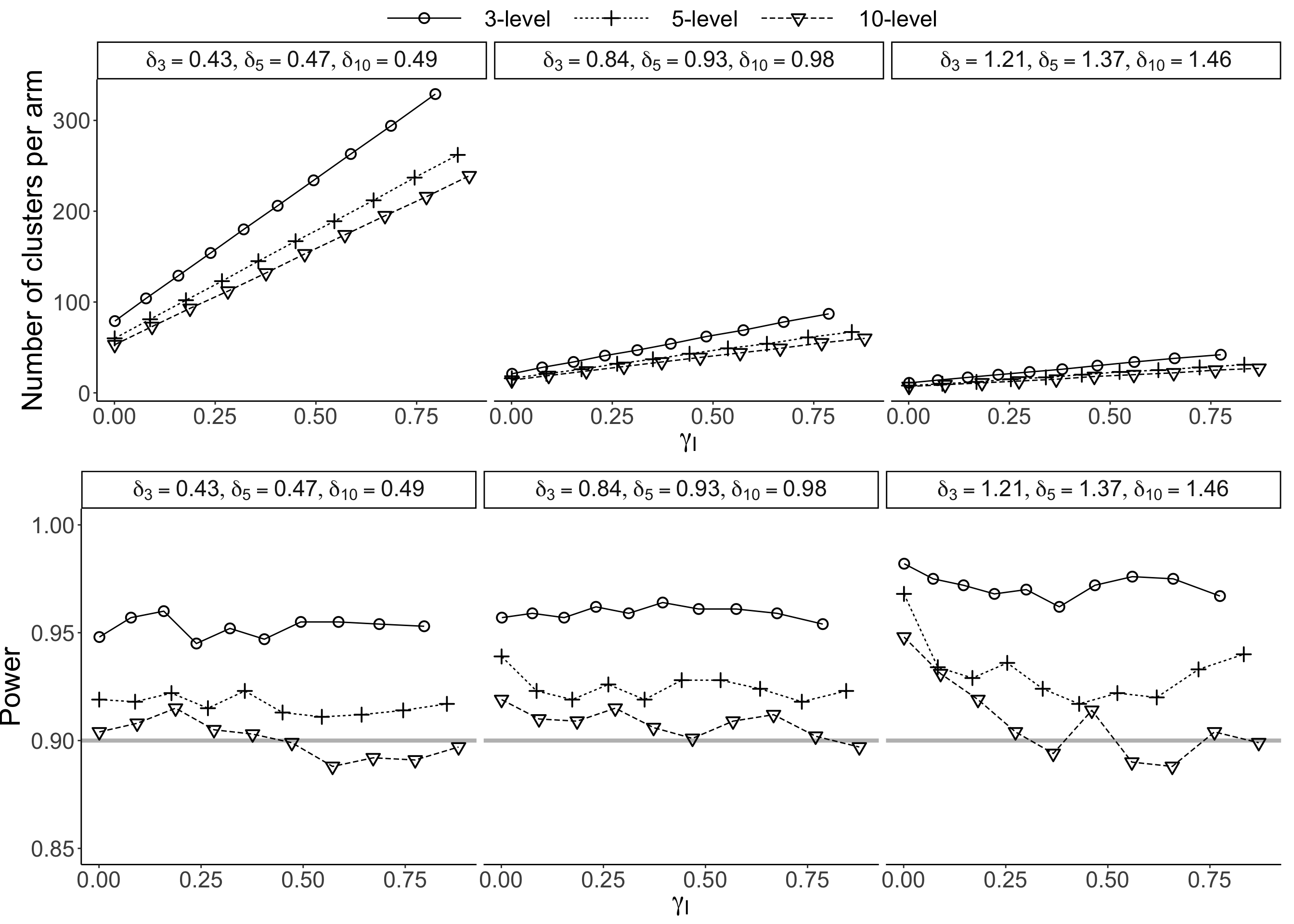} \label{fig:size_powerk5ord}}
     \quad
     \subfloat[Cluster sizes = 50]{\includegraphics[width=0.8\textwidth]{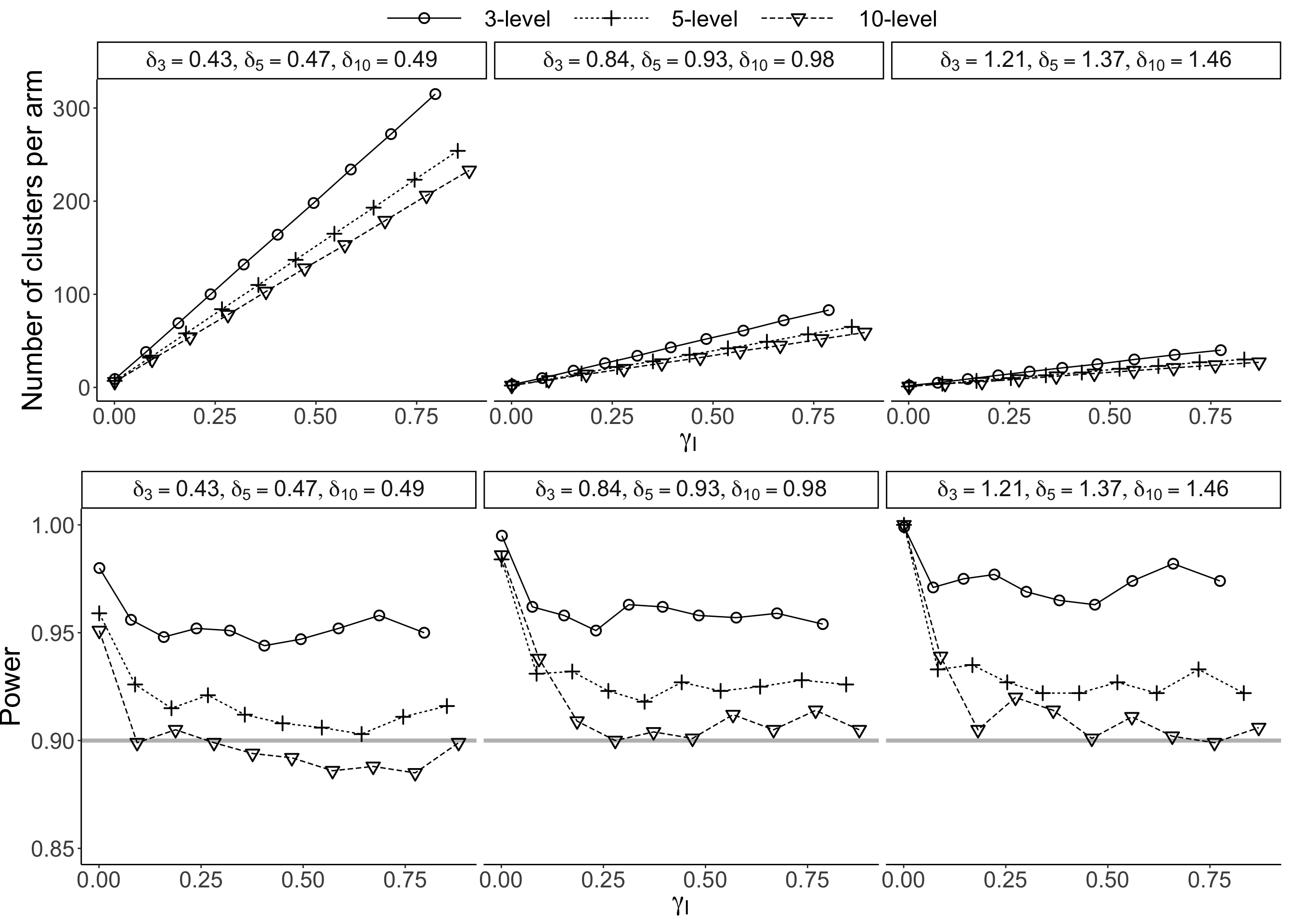} \label{fig:size_powerk50ord} }
     \caption[]{Number of clusters per arm and observed power for ordinal data with predetermined cluster sizes of 5 and 50 with the number of clusters selected as that needed to achieve 90\% power. $\delta_3$, $\delta_5$, $\delta_{10}$ are the log ORs of the 3-level, 5-level, and 10-level ordinal outcomes, respectively.}
     \label{fig:size_power_ord}
\end{figure*}

We also evaluated the performance of our sample size calculations for ordinal data. We generated clustered data of 3-level, 5-level, and 10-level ordinal variables by discretizing $X_{0ij}$ and $Y_{0ts}$ with cut-offs at quantiles of a standard normal distribution (i.e., using the 1/3 and 2/3 quantiles for 3 levels; the 0.2, 0.4, 0.6, 0.8 quantiles for 5 levels; and the 0.1, 0.2, ..., 0.8, 0.9 quantiles for 10 levels). To calculate sample sizes for each ordinal variable using formula (\ref{noc}), we derived $\bar{\pi}_l$ based on the known distributions and empirically computed values for $\gamma_I$ and $\delta$ based on the specific data generation scenario. Empirically computed values for $\gamma_I$ and $\delta$ were obtained by generating a million clusters and 100 observations per cluster. The empirical values of $\gamma_I$ and $\delta$ of the 3-level ordinal outcome were slightly smaller than those of the other two ordinal outcomes. In summary, our sample size calculations had good power for ordinal data in most scenarios. The calculated number of clusters for the three ordinal variables all increased as the $\gamma_I$ increased (Figure \ref{fig:size_power_ord}). The numbers of clusters calculated for the three ordinal variables, in descending order, were as follows: 3-level $>$ 5-level $>$ 10-level. The powers were in the same order from largest to smallest. The power of the 3-level ordinal variable was around 0.95, indicating slight overestimation in the number of clusters for this variable. Whitehead's sample size calculations for individual RCTs have the same issue when the number of categories is very small (Web Figure 4). This is because of an approximation used to derive Whitehead's formula. We analytically show that this approximation becomes better, and hence the overestimation of the sample size in Whitehead's formula decreases as the number of categories increases (see Supplementary Materials Appendix A for details). 

Our sample size calculations can also be applied to binary outcomes, which can be treated as ordinal with two categories. We show via simulation (Web Figure 6) that sample sizes obtained by our sample size calculations are very similar to those obtained by commonly used calculations for binary outcomes \citep{hayes2009cluster}. Those commonly used calculations treat $1+\rho_I(k-1)$ as the DE to inflate the sample size of an individual RCT \citep{hayes2009cluster}. Notably, the ICC $\rho_I$ for binary outcomes is equal to the rank ICC for binary outcomes. 

\section{Applications}
\label{s:applications}
\subsection{An individual RCT for a skewed outcome with detection limits}
Researchers are interested in understanding the effect of an experimental statin treatment (HMG-CoA-reductase inhibitor) on inflammatory markers in people living with HIV (PWH) in Nigeria \citep{soko2016}. The specific biomarkers of interest are interleukin 6 (IL-6), high sensitivity C-reactive protein (hsCRP), and soluble CD14 (SCD14) \citep{hunt2012, kuller2008}. Each of these biomarkers has been associated with diabetes, cardiovascular disease, and all-cause mortality. Preliminary data for these variables in the Nigerian study population are not available; however, the literature contains measures of these variables in different populations. Notably, each of the three biomarkers tends to be right-skewed to varying extents, suggesting different transformations might be needed for each of the outcomes. In addition, IL-6 is subject to a detection limit, with values below a certain level, denoted DL, simply recorded as $<$DL. 

In settings such as this with little preliminary data available on skewed outcomes, sample sizes are often computed to detect a mean difference on a \lq\lq suitably" transformed scale in terms of standard deviations (SD). For example, using standard sample size calculations for a two-sample t-test in a 1:1 randomized trial, total sample sizes of 34, 128, and 506 would be needed to have 80\% power to detect mean differences of 1, 0.5, and 0.25 SDs, respectively, on the suitably transformed scale (assuming a two-sided significance level of 0.05 for this and all other calculations). Although often done in practice, there are problems with this calculation. First, the implicit assumption that one will be able to correctly select the suitable transformation at the analysis stage may be strong. Second, translation of this effect size to the original scale may be needed to interpret the sample size calculations, which requires hypothesizing a transformation at the design stage. For example, based on the data presented in Tian et al. \citep{tian2020}, a 1 SD difference on the log-transformed scale suggests a difference in median IL-6 on the original scale of approximately 3 versus 12 pg/ml, whereas a 1 SD difference on the square root transformed scale suggests a difference in median IL-6 of approximately 3 versus 7 pg/ml. 

In contrast, our sample size calculations do not require specifying a transformation of the data. Our calculations are formulated with odds ratios or probabilistic indices. With a continuous outcome, the proportional odds model implicitly assumes a common odds ratio across all order-maintaining dichotomizations of the response variable. Thus, the odds ratio could be elicited by considering an arbitrary dichotomization of the outcome (e.g., dichotomizing at the median). To have 80\% power to detect odds ratios of 3, 2, and 1.5, total sample sizes of 80, 198, and 574, respectively, would be required.  In terms of probabilistic indexes, to have 80\% power to detect $\theta$ of 0.65, 0.6, and 0.55, total sample sizes of 110, 254, and 1042, respectively, would be required. These sample sizes are larger than those of the previous paragraph, because differences (after a suitable transformation) on the logistic latent scale of 1, 0.5, and 0.25 SDs correspond to odds ratios of approximately 6.13, 2.48, and 1.57, respectively, because OR$=\exp(\text{SD}\times\pi/\sqrt{3})$. Using these OR, our sample size formula suggests that total sample sizes of 30, 116, and 460, respectively, would be required.

\subsection{A cluster randomized trial with a skewed continuous outcome}
The HoPS+ study conducted a cluster randomized trial to measure the impact of a multi-component intervention on adherence to ART of pregnant women living with HIV in $\text{Zamb}\acute{\text{e}}\text{zia}$ Province, Mozambique \citep{audet2018, audet2024}. The primary outcome was adherence to ART, quantified as the proportion of days the medications were taken within 1 year. This measure of adherence is pseudo-continuous ranging from 0 to 1 with a left-skewed distribution; many participants were highly adherent, while others had moderate to poor adherence. Sample size calculations for this trial were based on a simplified binary outcome (i.e., retention at 6 months) and an assumed ICC of 0.07 for this binary outcome. The study design anticipated having approximately 85\% power to detect an improvement in 6-month retention from 31\% to 48\% (equivalent to OR $=2.05$) with a two-sided significance level of 0.05 and cluster sizes of 45. The number of clusters (clinics) under this design was 24 with 12 per arm. This sample size calculation based on a simplified binary outcome, while fairly standard, was conservative, likely resulting in a larger sample size than needed. 

We recalculated the sample size with our sample size calculations but using the primary outcome without dichotomization. Since the primary outcome ranged from 0 to 1 with a left-skewed distribution and 366 possible values, the outcome can be treated either as a continuous or an ordinal variable with our sample size calculations. If the primary outcome is treated as ordinal, the proportion in each ordered category must be estimated to calculate the sample size. Now the trial is over, supposing we do not have access to the data, which is common in practice, these proportions can be roughly estimated post-hoc using published data \citep{tu2024between}. In contrast, if the primary outcome is treated as continuous, the calculation is much simpler because it does not require a preliminary estimate of the distribution. It turns out that whether the primary outcome was considered either continuous or ordinal with proportions post-hoc estimated from the published trial, the calculated number of clinics (i.e., clusters) remained the same. This is expected because the outcome is roughly continuous with 366 possible values between 0 and 1, and no single proportion was very large. Therefore, we consider the outcomes to be continuous in all of the following calculations in this subsection. 

Under the same setting as the original design (power of 85\%, type I error rate of 0.05, cluster sizes of 45, rank ICC of 0.07, and OR $= \exp(\delta)=2.05$), our sample size calculations yielded a calculated number of clinics of 10 per arm, which is smaller than that of the original design (12 per arm). Alternatively, with 12 clusters per arm, to achieve the same power with the continuous outcome, we would only have needed clusters of size 21. 
 
\begin{figure*}
     \centering
     \subfloat[Total number of clinics calculated with predetermined cluster sizes]{\label{fig:phq1}  \includegraphics[width=0.5\textwidth]{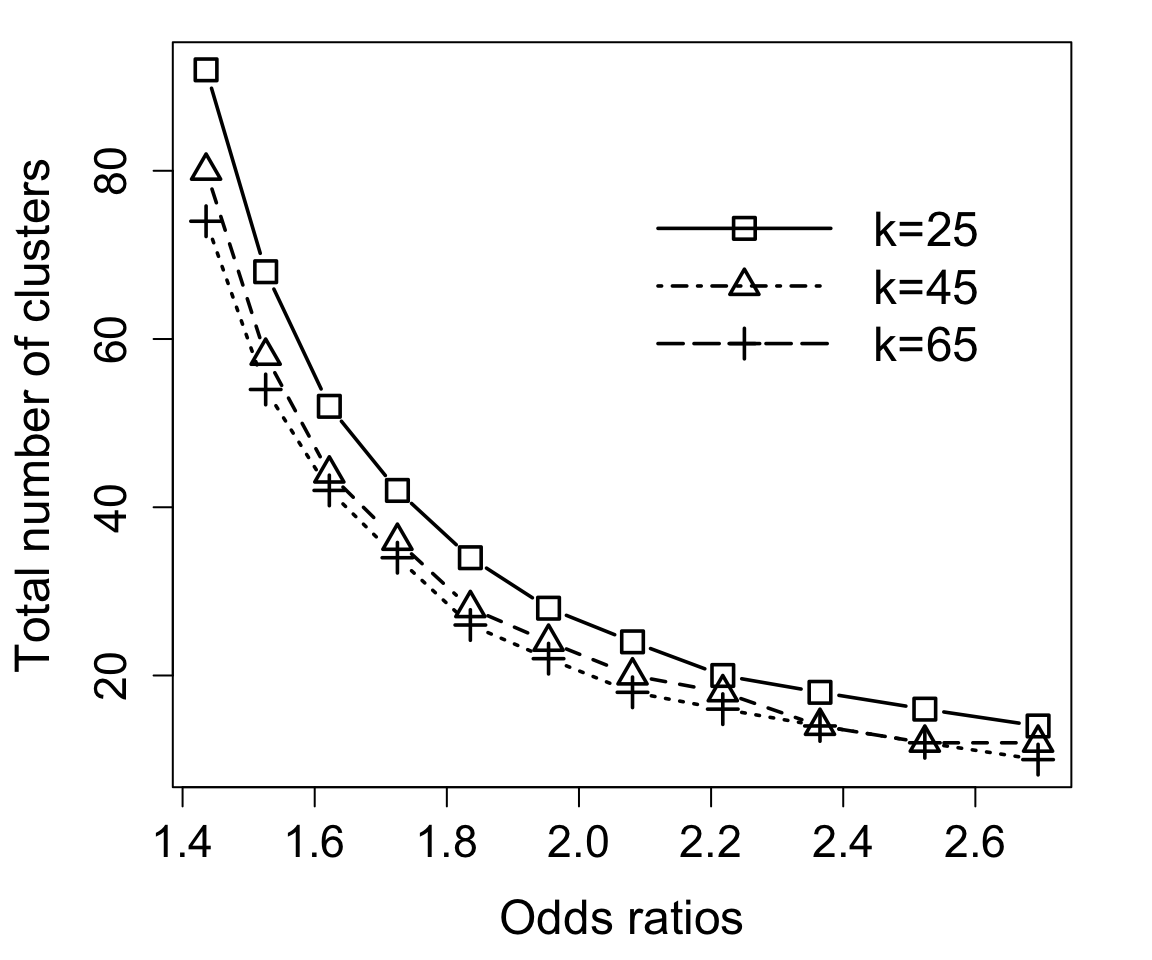}}
     \quad
     \subfloat[Cluster sizes calculated with predetermined total numbers of clinics]{\label{fig:phq2} \includegraphics[width=0.5\textwidth]{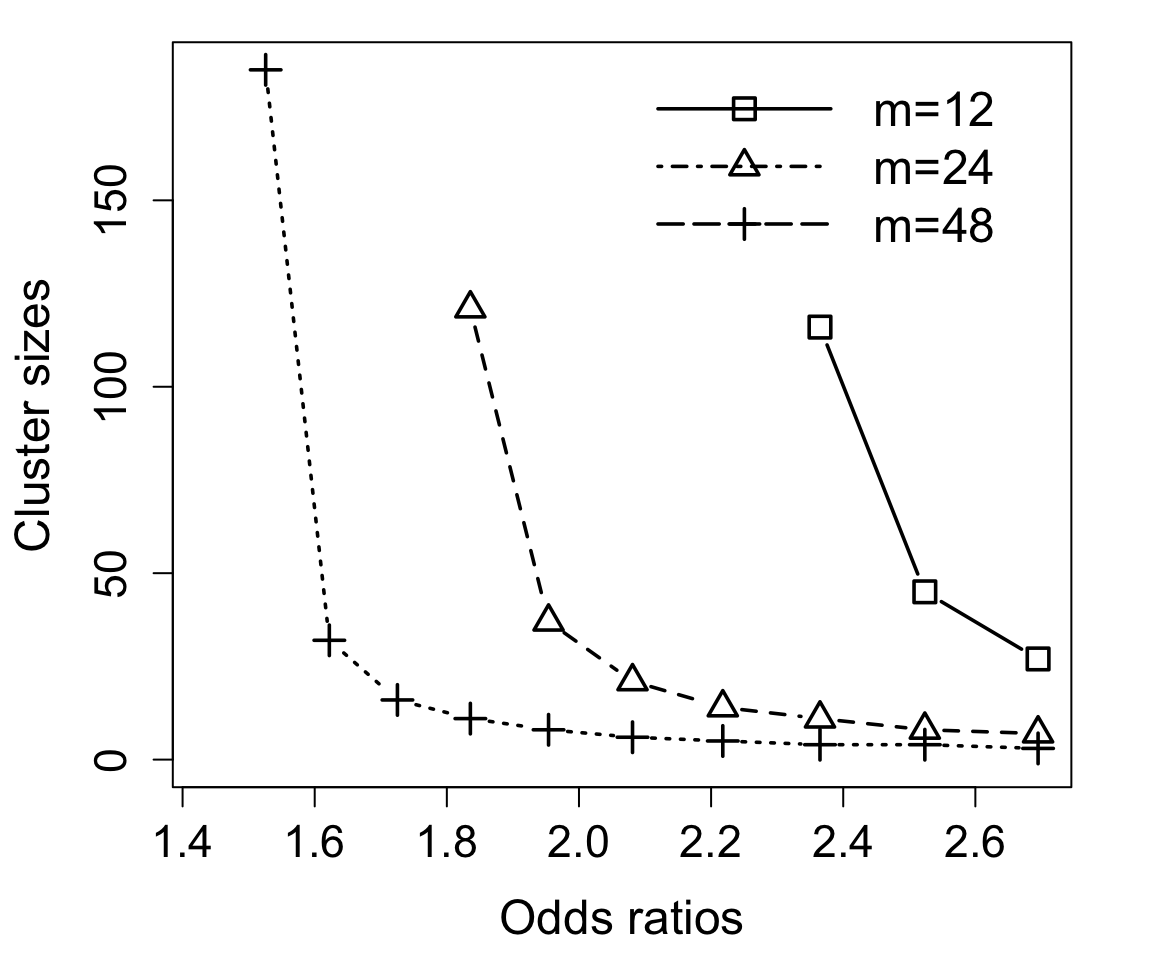}}
     \caption[Results for the HoPS+ study example]{Numbers of clinics (clusters) and individuals per clinic needed to have power of 0.85 to detect the given odds ratios in the HoPS+ study example. The cluster size is denoted by $k$ and the total number of clinics is denoted by $m$. The two-sided significance level was set to 0.05, and the rank ICC was set to 0.074.}
     \label{fig:phq}
\end{figure*}

The rank ICC of 12-month adherence for women in the HoPS+ study was 0.074 \citep{tu2024between}, close to the assumed ICC of 0.07. Using the rank ICC of 0.074, we calculated the sample size across different ORs (Figure \ref{fig:phq}). As the OR increased, the calculated sample size had an initial rapid decrease followed by a gradual decrease. Increasing the numbers in each cluster helped to reduce the required number of clusters, but at some point the benefits became incremental. In addition, a limited predetermined number of clusters may hinder the detection of small treatment effects, even in cases when the rank ICC is small (Figure \ref{fig:phq2}). 

\subsection{A non-inferiority cluster randomized clinical trial with an ordinal outcome}
In the BRIDGE study, a non-inferiority RCT was designed to understand if task-shifting childhood epilepsy treatment by trained community health workers was non-inferior to enhanced physician care in reducing seizures \citep{aliyu2019}. This trial identified children with untreated epilepsy from communities served by community-based primary healthcare centers (PHCs) in northern Nigeria and recruited these children into the cluster RCT. The intervention was task-shifted epilepsy care by trained community health workers (TSC). The control was ``enhanced usual physician care'' (EUC, referral to a physician plus primary care by epilepsy-trained community healthcare workers helping study subjects navigate the healthcare system). The study had general inclusion criteria which included children with all seizure types except for infantile spasms. The primary outcome was whether the child had been seizure-free (yes/no) for 6 months or more at the 24-month follow-up visit. The definition of this binary outcome is standard in this setting where children with a wide variety of seizure types (convulsive and non-convulsive) and seizure frequencies were included. Counts for some non-convulsive seizure types (e.g., absence seizures) can only be estimated, whereas seizure-freedom versus no seizure-freedom can be accurately determined. However, convulsive seizure types can be accurately counted. The one-sided null hypothesis was that the seizure-free rate of TSC patients (intervention) was inferior to that of EUC patients (control) by $\geq$10\% (equivalent to the log OR $\geq1.5$). The one-sided significance level and required power were set to 0.05 and 0.8, respectively, and the ICC was assumed to be 0.05. With a predetermined number of clusters at 30 per arm, the cluster size was calculated to be 19.

\begin{figure*}
     \centering
     \subfloat[Histogram of convulsive seizure counts between 18- and 24-month visits by the two arms]{ \label{fig:seizurehist}  \includegraphics[width=0.6\textwidth]{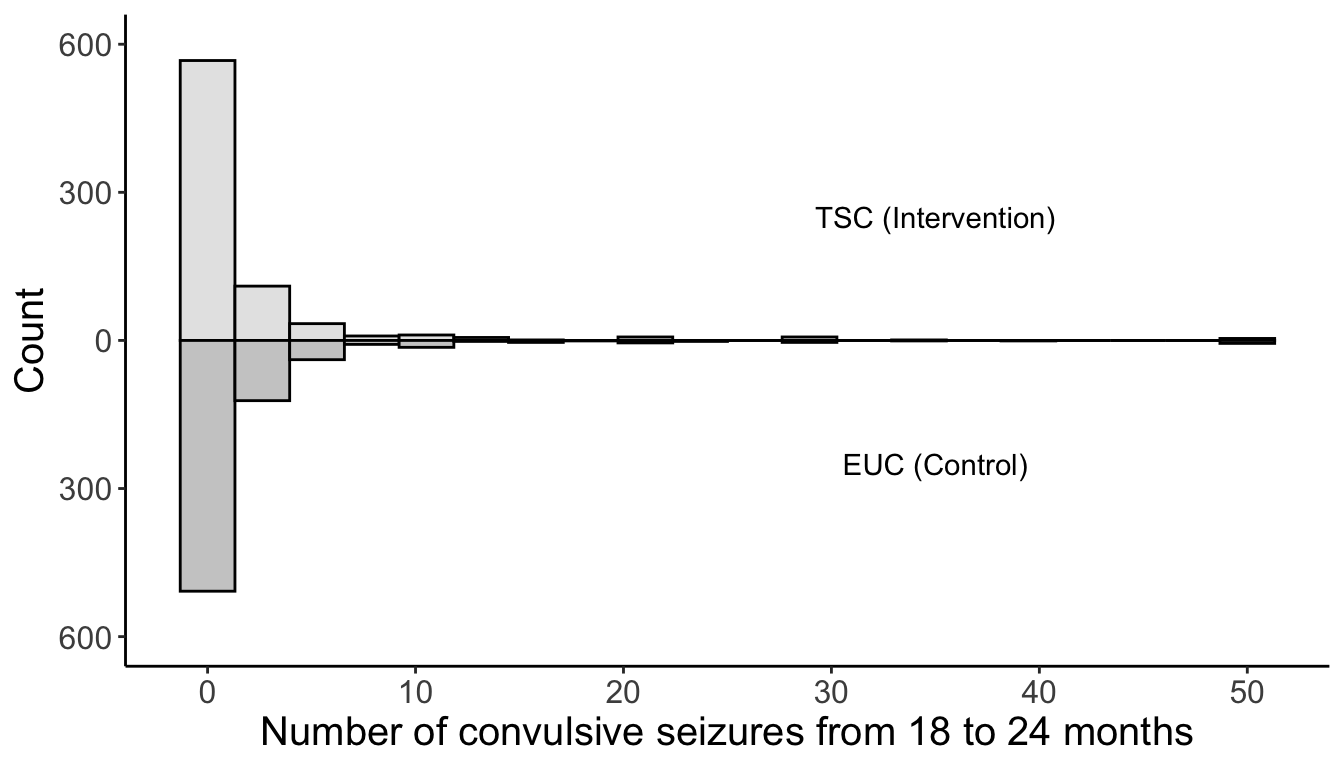}}
     \quad
     \subfloat[Total number of clusters calculated with predetermined cluster sizes]{\label{fig:seizure1}  \includegraphics[width=1\textwidth]{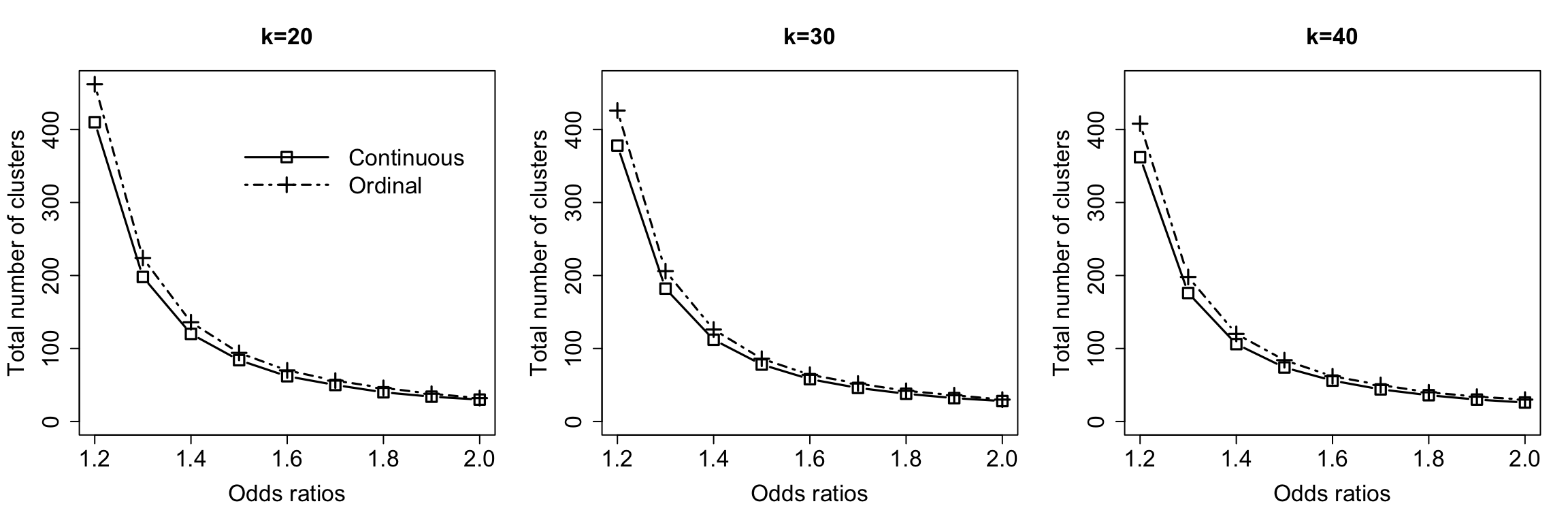}}
     \quad
     \subfloat[Cluster sizes calculated with predetermined total numbers of clusters]{ \label{fig:seizure2}   \includegraphics[width=1\textwidth]{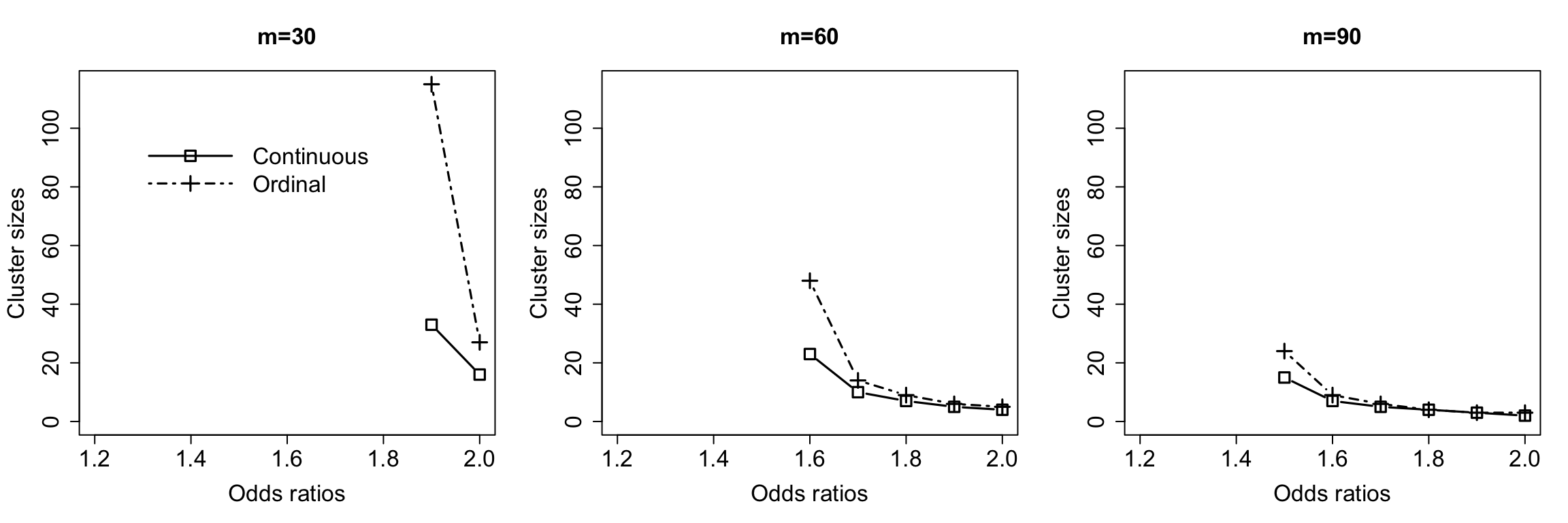}}
     \caption[Results for the BRIDGE trial example]{Histogram of convulsive seizure counts between 18- and 24-month visits, numbers of clusters and individuals per cluster needed to have the power of 0.8 to detect the given odds ratios in the BRIDGE study example. The cluster size is denoted by $k$ and the total number of clusters is denoted by $m$. ``Continuous'' represents treating the outcome as continuous and ``Ordinal'' represents treating the outcome as ordinal. The one-sided significance level was set to 0.05, and the rank ICC was set to 0.14.}
     \label{fig:seizure}
\end{figure*}

There may be interest in performing a new study to examine interventions on children with convulsive seizures, which can be accurately counted. In this more homogeneous subpopulation, the number of seizures in the past 6 months at the 24-month follow-up visit is a scientifically meaningful response variable, resulting in higher power than a dichotomized (0 versus $\geq$1) response variable. We consider the same one-sided hypothesis in the design of a new cluster RCT among children with convulsive seizures, with the primary outcome being the number of seizures from months 18 to 24. The BRIDGE study data among the subset of children with convulsive seizures can be used as preliminary data. A histogram of convulsive seizure counts between months 18 and 24 in the BRIDGE trial is given in Figure \ref{fig:seizurehist}. 

Since the new primary outcome is an irregularly distributed count variable, it is reasonable to treat it as an ordinal variable. The proportion of responses for each ordered category can be easily estimated from the BRIDGE study data. One could alternatively treat the number of convulsive seizures as a continuous variable for sample size calculations, where the proportion of each outcome is 1 over the total sample size (even though in the preliminary data, it only took integer values from 0 to 50). We compared the sample sizes when treating the new outcome as ordinal and estimating $\overline{\pi}_l$ using the observed proportions in the BRIDGE trial (i.e., formula (\ref{noc})) versus treating the outcome as continuous (i.e., formula (\ref{ncont})). The one-sided significance level and required power were set to 0.05 and 0.8, respectively. The rank ICC was 0.14, estimated from the BRIDGE study. The results are shown in Figures \ref{fig:seizure1} and \ref{fig:seizure2}. The calculation treating the outcome as ordinal yielded larger sample sizes than the calculation treating the outcome as continuous. This difference is expected as a large proportion of participants in the BRIDGE study had no seizures, i.e., $\overline{\pi}_0$ is large. Therefore, for the proposed study, it would be preferable to use formula (\ref{noc}) to compute sample sizes. These calculations suggest that to rule out non-inferiority (OR $\geq$ 1.5) we would need $45$ clusters per arm each with $k=24$ or $42$ clusters per arm each with $k=40$. These sample sizes are larger than those in the original trial because the rank ICC for the count outcome (estimated from the original trial) is higher than what was assumed for the original trial for the binary seizure-free outcome (0.14 versus 0.05). 

\section{Discussion}
\label{s:disucssion}
In this paper, we propose unified and simple sample size calculations for individual or cluster RCTs with skewed continuous or ordinal outcomes. We show that the sample size calculations introduced by Whitehead \citep{whitehead1993} for individual RCTs with ordinal outcomes can be easily extended to continuous outcomes. We extended these calculations to cluster RCTs by inflating the sample size for an adequately powered individual RCT for an ordinal outcome with a design effect that incorporates the rank ICC. In simulations, our sample size calculations achieved power at approximately the desired level in nearly all scenarios we considered for both skewed continuous and ordinal data. With continuous outcomes, our sample size calculations yield sample sizes that are very close to the sample sizes obtained by more complex calculations based on Wilcoxon rank sum tests. Our calculations are useful for individual RCTs with skewed continuous outcomes because they make minimal assumptions on the unknown distribution of the outcome. 

Our sample size calculations in (\ref{noc}) and (\ref{ncont}) use the rank ICC of the entire population, although the rank ICC may differ between study arms. To address this, one might use preliminary data to separately estimate the rank ICCs of the two arms, and then calculate the sample size of each arm by separately inflating sample sizes using DEs based on their respective rank ICCs. In practice, however, it is often difficult to obtain precise rank ICC estimates for both arms given limited preliminary data. 

Our sample size calculations have some limitations. They require specifying the effect size using an odds ratio, which may be unnatural for continuous outcomes, although well-defined (see Section \ref{s:review}). For some people, specifying the effect size using the probabilistic index, which can easily be converted to an odds ratio, may be preferable. Another limitation, which is shared by nearly all sample size calculations, is that they are for unadjusted analyses, whereas in practice we often adjust for covariates as it is well known that power can increase with adjustment for carefully chosen baseline covariates \citep{Kahan2014}. Hence, power will likely be higher than anticipated if one adjusts for covariates. Note that PO and cluster PO models allow easy inclusion of covariates \citep{tian2023}. For ordinal data with a very small number of ordered categories, Whitehead's formula, and thus our sample size calculations, may overestimate the required sample size. In addition, our calculations assume equal cluster sizes. In simulations, we saw that if cluster size imbalance is not extreme, our calculations remain applicable; otherwise, they might yield an underestimate. Future work could consider developing calibration approaches for ordinal outcomes with very few ordered categories and improving the calculations to accommodate unbalanced cluster sizes.

For most researchers, current practice with skewed continuous outcomes involves either 1) treating the outcome as dichotomous and calculating sample sizes using known formulas for RCTs with binary outcomes, or 2) assuming the outcome will be approximately normally distributed after some transformation and calculating sample sizes using known formulas for RCTs to detect a difference of means on the transformed scale. The first approach is usually justified in that the sample size computed for a binary outcome should result in greater-than-anticipated power when it is analyzed as a continuous variable; and that recruiting more than the needed number of study participants is thought of as a lesser problem than recruiting fewer participants than needed. However, recruiting more participants than needed is not completely benign: under such a strategy, study costs are higher than needed, study feasibility decreases, and more participants than needed could be assigned to a study arm that is potentially inferior to their health. The second approach is often justified because researchers are confident they will be able to find a meaningful transformation once they obtain the data. However, proper transformations are often difficult to select, results may be sensitive to the choice of transformation, and interpretation on the transformed scale may be challenging. In essence, our recommendation is to prespecify at the design phase that these skewed outcomes will be analyzed in a rank-based manner and to power the trial in a manner consistent with this rank-based analysis. Our sample size calculations allow investigators to power their study in such a manner.

\section*{Competing interests}
No competing interest is declared.


\section*{Acknowledgments}
We would like to thank the study investigators for providing data used in our example applications. This study was supported in part by funding from the National Institutes of Health (R01AI093234; R01MH113478 for HoPS+; and R01NS113171 for BRIDGE).

\bibliographystyle{biom} 
\bibliography{reference.bib}

\begin{thebibliography}{}

\bibitem[\protect\citeauthoryear{Aliyu, Abdullahi, Iliyasu, Salihu, Adamu, Sabo, et~al\mbox{.}}{Aliyu et~al.}{2019}]{aliyu2019}
Aliyu, M.~H., Abdullahi, A.~T., Iliyasu, Z., Salihu, A.~S., Adamu, H., Sabo, U., et~al. (2019).
\newblock Bridging the childhood epilepsy treatment gap in northern {Nigeria} ({BRIDGE}): Rationale and design of pre-clinical trial studies.
\newblock {\em Contemp Clin Trials Commun} {\bf 15,} 100362.

\bibitem[\protect\citeauthoryear{Audet, Graves, Barreto, De~Schacht, Gong, Shepherd, et~al\mbox{.}}{Audet et~al.}{2018}]{audet2018}
Audet, C.~M., Graves, E., Barreto, E., De~Schacht, C., Gong, W., Shepherd, B.~E., et~al. (2018).
\newblock Partners-based {HIV} treatment for seroconcordant couples attending antenatal and postnatal care in rural {Mozambique}: A cluster randomized trial protocol.
\newblock {\em Contemp Clin Trials Commun} {\bf 71,} 63--69.

\bibitem[\protect\citeauthoryear{Audet, Graves, Shepherd, Prigmore, Brooks, Emilio, Matino, Paulo, Diemer, Frisby, Sack, Aboobacar, Barreto, Van~Rompaey, and De~Shacht}{Audet et~al.}{2024}]{audet2024}
Audet, C.~M., Graves, E., Shepherd, B.~E., Prigmore, H.~L., Brooks, H.~L., Emilio, A., Matino, A., Paulo, P., Diemer, M.~A., Frisby, M., Sack, D.~E., Aboobacar, A., Barreto, E., Van~Rompaey, S., and De~Shacht, C. (2024).
\newblock Partner-based hiv treatment for seroconcordant couples attending antenatal and postnatal care in rural {Mozambique}: a cluster randomized controlled trial.
\newblock {\em Journal of Acquired Immune Deficiency Syndromes} page (in press).

\bibitem[\protect\citeauthoryear{Bross}{Bross}{1958}]{Bross1958}
Bross, I. D.~J. (1958).
\newblock How to use ridit analysis.
\newblock {\em Biometrics} {\bf 14,} 18--38.

\bibitem[\protect\citeauthoryear{Campbell and Walters}{Campbell and Walters}{2014}]{campbell2014}
Campbell, M. and Walters, S. (2014).
\newblock {\em How to Design, Analyse and Report Cluster Randomised Trials in Medicine and Health Related Research}.
\newblock Wiley, Chichester.

\bibitem[\protect\citeauthoryear{De~Neve, Thas, and Gerds}{De~Neve et~al.}{2019}]{de_neve2019}
De~Neve, J.~D., Thas, O., and Gerds, T.~A. (2019).
\newblock Semiparametric linear transformation models: Effect measures, estimators, and applications.
\newblock {\em Statistics in Medicine} {\bf 38,} 1484--1501.

\bibitem[\protect\citeauthoryear{Denham}{Denham}{2016}]{denham2016}
Denham, B.~E. (2016).
\newblock {\em Categorical Statistics for Communication Research}.
\newblock John Wiley and Sons, Ltd.

\bibitem[\protect\citeauthoryear{Donner, Birkett, and Buck}{Donner et~al.}{1981}]{donner1981}
Donner, A., Birkett, N., and Buck, C. (1981).
\newblock Randomization by cluster-sample size requirements and analysis.
\newblock {\em American Journal of Epidemiology} {\bf 114,} 906--914.

\bibitem[\protect\citeauthoryear{Fisher}{Fisher}{1925}]{fisher1925}
Fisher, R.~A. (1925).
\newblock {\em Statistical methods for research workers.}
\newblock Edinburgh Oliver \& Boyd.

\bibitem[\protect\citeauthoryear{Hayes and Moulton}{Hayes and Moulton}{2009}]{hayes2009cluster}
Hayes, R.~J. and Moulton, L.~H. (2009).
\newblock {\em Cluster Randomised Trials}.
\newblock Chapman \& Hall/CRC.

\bibitem[\protect\citeauthoryear{Heagerty and Zeger}{Heagerty and Zeger}{1996}]{heagerty1996}
Heagerty, P.~J. and Zeger, S.~L. (1996).
\newblock Marginal regression models for clustered ordinal measurements.
\newblock {\em Journal of the American Statistical Association} {\bf 91,} 1024--1036.

\bibitem[\protect\citeauthoryear{Hollander and Wolfe}{Hollander and Wolfe}{1999}]{hollander1999}
Hollander, M. and Wolfe, D.~A. (1999).
\newblock {\em Nonparametric Statistical Methods}.
\newblock New York: John Wiley \& Sons, Inc.

\bibitem[\protect\citeauthoryear{Hunt}{Hunt}{2012}]{hunt2012}
Hunt, P. (2012).
\newblock Hiv and inflammation: mechanisms and consequences.
\newblock {\em Current HIV/AIDS Reports} {\bf 9,} 139--147.

\bibitem[\protect\citeauthoryear{Kahan, Jairath, Dor{\'e}, and Morris}{Kahan et~al.}{2014}]{Kahan2014}
Kahan, B.~C., Jairath, V., Dor{\'e}, C.~J., and Morris, T.~P. (2014).
\newblock The risks and rewards of covariate adjustment in randomized trials: an assessment of 12 outcomes from 8 studies.
\newblock {\em Trials} {\bf 15,} 139.

\bibitem[\protect\citeauthoryear{Kish}{Kish}{1965}]{kish1965}
Kish, L. (1965).
\newblock {\em Survey Sampling}.
\newblock New York: John Wiley \& Sons, Inc.

\bibitem[\protect\citeauthoryear{Kish}{Kish}{1987}]{kish1987}
Kish, L. (1987).
\newblock Weighting in deft.
\newblock {\em The Survey Statistician} .

\bibitem[\protect\citeauthoryear{Kuller, Tracy, Belloso, and et~al.}{Kuller et~al.}{2008}]{kuller2008}
Kuller, L., Tracy, R., Belloso, W., and et~al. (2008).
\newblock Inflammatory and coagulation biomarkers, and mortality in patients with hiv infection.
\newblock {\em PLOS Medicine} {\bf 5,} e203.

\bibitem[\protect\citeauthoryear{Lehmann}{Lehmann}{1975}]{lehmann1975}
Lehmann, E.~L. (1975).
\newblock {\em Nonparametrics: Statistical Methods Based on Ranks}.
\newblock San Francisco: Holden-Day.

\bibitem[\protect\citeauthoryear{Liu, Shepherd, Li, and Harrell~Jr.}{Liu et~al.}{2017}]{liu2017}
Liu, Q., Shepherd, B.~E., Li, C., and Harrell~Jr., F.~E. (2017).
\newblock Modeling continuous response variables using ordinal regression.
\newblock {\em Statistics in Medicine} {\bf 36,} 4316--4335.

\bibitem[\protect\citeauthoryear{Mann and Whitney}{Mann and Whitney}{1947}]{mann1947}
Mann, H.~B. and Whitney, D.~R. (1947).
\newblock On a test of whether one of two random variables is stochastically larger than the other.
\newblock {\em Annals of Mathematical Statistics} {\bf 18,} 50--60.

\bibitem[\protect\citeauthoryear{McCullagh}{McCullagh}{1980}]{mccullagh1980}
McCullagh, P. (1980).
\newblock Regression models for ordinal data.
\newblock {\em Journal of the Royal Statistical Society, Series B} {\bf 43,} 109--142.

\bibitem[\protect\citeauthoryear{Parsons, Edmondson, and Gilmour}{Parsons et~al.}{2006}]{parsons2006}
Parsons, N.~R., Edmondson, R., and Gilmour, S. (2006).
\newblock A generalized estimating equation method for fitting autocorrelated ordinal score data with an application in horticultural research.
\newblock {\em Journal of the Royal Statistical Society, Series C (Applied Statistics)} {\bf 55,} 507--524.

\bibitem[\protect\citeauthoryear{Pearson}{Pearson}{1907}]{Pearson1907}
Pearson, K.~G. (1907).
\newblock {\em On Further Methods of Determining Correlation}.
\newblock Cambridge University Press, Cambridge.

\bibitem[\protect\citeauthoryear{Rosner and Glynn}{Rosner and Glynn}{2011}]{rosner2011}
Rosner, B. and Glynn, R. (2011).
\newblock Power and sample size estimation for the clustered wilcoxon test.
\newblock {\em Biometrics} {\bf 67,} 646–653.

\bibitem[\protect\citeauthoryear{Rosner and Glynn}{Rosner and Glynn}{2009}]{rosner2009}
Rosner, B. and Glynn, R.~J. (2009).
\newblock Power and sample size estimation for the wilcoxon rank sum test with application to comparisons of c statistics from alternative prediction models.
\newblock {\em Biometrics} {\bf 65,} 188--197.

\bibitem[\protect\citeauthoryear{Rosner, Glynn, and Lee}{Rosner et~al.}{2003}]{rosner2003}
Rosner, B., Glynn, R.~J., and Lee, M.-L.~T. (2003).
\newblock Incorporation of clustering effects for the wilcoxon rank sum test: A large sample approach.
\newblock {\em Biometrics} {\bf 59,} 1089--1098.

\bibitem[\protect\citeauthoryear{Rosner, Glynn, and Lee}{Rosner et~al.}{2006}]{rosner2006}
Rosner, B., Glynn, R.~J., and Lee, M.-L.~T. (2006).
\newblock Extension of the rank sum test for clustered data: two-group comparisons with group membership defined at the subunit level.
\newblock {\em Biometrics} {\bf 62,} 1251--1259.

\bibitem[\protect\citeauthoryear{Rutterford, Copas, and Eldridge}{Rutterford et~al.}{2015}]{rutterford2015}
Rutterford, C., Copas, A., and Eldridge, S. (2015).
\newblock Methods for sample size determination in cluster randomized trials.
\newblock {\em International Journal of Epidemiology} {\bf 44,} 1051--1067.

\bibitem[\protect\citeauthoryear{Siegel}{Siegel}{1956}]{siegel1956}
Siegel, S. (1956).
\newblock {\em Nonparametric Statistics for the Behavioural Sciences}.
\newblock New York: McGraw-Hill.

\bibitem[\protect\citeauthoryear{Soko, Masimirembwa, and Dandara}{Soko et~al.}{2016}]{soko2016}
Soko, N., Masimirembwa, C., and Dandara, C. (2016).
\newblock Pharmacogenomics of rosuvastatin: A glocal (global+local) african perspective and expert review on a statin drug.
\newblock {\em OMICS} {\bf 20,} 498--509.

\bibitem[\protect\citeauthoryear{Tian, Hothorn, Li, Harrell, and Shepherd}{Tian et~al.}{2020}]{tian2020}
Tian, Y., Hothorn, T., Li, C., Harrell, F.~J., and Shepherd, B.~E. (2020).
\newblock An empirical comparison of two novel transformation models.
\newblock {\em Stat Med} {\bf 39,} 562--576.

\bibitem[\protect\citeauthoryear{Tian, Li, Tu, James, Harrell, and Shepherd}{Tian et~al.}{2024}]{tian2024addressing}
Tian, Y., Li, C., Tu, S., James, N.~T., Harrell, F.~E., and Shepherd, B.~E. (2024).
\newblock Addressing multiple detection limits with semiparametric cumulative probability models.
\newblock {\em Journal of the American Statistical Association} {\bf 119,} 864--874.

\bibitem[\protect\citeauthoryear{Tian, Shepherd, Li, Zeng, and Schildcrout}{Tian et~al.}{2023}]{tian2023}
Tian, Y., Shepherd, B.~E., Li, C., Zeng, D., and Schildcrout, J.~S. (2023).
\newblock Analyzing clustered continuous response variables with ordinal regression models.
\newblock {\em Biometrics} {\bf 79,} 3764--3777.

\bibitem[\protect\citeauthoryear{Tu, Li, and Shepherd}{Tu et~al.}{2024}]{tu2024between}
Tu, S., Li, C., and Shepherd, B.~E. (2024).
\newblock Between- and within-cluster {Spearman} rank correlations.
\newblock {\em arXiv preprint arXiv:2402.11341} .

\bibitem[\protect\citeauthoryear{Tu, Li, Zeng, and Shepherd}{Tu et~al.}{2023}]{tu2023}
Tu, S., Li, C., Zeng, D., and Shepherd, B.~E. (2023).
\newblock Rank intraclass correlation for clustered data.
\newblock {\em Statistics in Medicine} {\bf 42,} 4333--4348.

\bibitem[\protect\citeauthoryear{Whitehead}{Whitehead}{1993}]{whitehead1993}
Whitehead, J. (1993).
\newblock Sample size calculations for ordered categorical data.
\newblock {\em Statistics in Medicine} {\bf 12,} 2257--2271.

\bibitem[\protect\citeauthoryear{Zeng and Lin}{Zeng and Lin}{2007}]{zeng2007}
Zeng, D. and Lin, D.~Y. (2007).
\newblock {Maximum Likelihood Estimation in Semiparametric Regression Models with Censored Data}.
\newblock {\em Journal of the Royal Statistical Society Series B: Statistical Methodology} {\bf 69,} 507--564.

\end{thebibliography}


\begin{thebibliography}{}

\bibitem[\protect\citeauthoryear{Chow, Shao, and Wang}{Chow et~al.}{2008}]{chow2008}
Chow, S.-C., Shao, J., and Wang, H. (2008).
\newblock {\em Sample Size Calculations in Clinical Research}.
\newblock CRC Press, Boca Raton, FL, 2nd edition.

\bibitem[\protect\citeauthoryear{De~Neve, Thas, and Gerds}{De~Neve et~al.}{2019}]{de_neve2019}
De~Neve, J.~D., Thas, O., and Gerds, T.~A. (2019).
\newblock Semiparametric linear transformation models: Effect measures, estimators, and applications.
\newblock {\em Statistics in Medicine} {\bf 38,} 1484--1501.

\bibitem[\protect\citeauthoryear{Fleiss, Levin, and Paik}{Fleiss et~al.}{2003}]{fleiss2003}
Fleiss, J.~L., Levin, B., and Paik, M.~C. (2003).
\newblock {\em Statistical Methods for Rates and Proportions}.
\newblock Wiley-Interscience, Hoboken, NJ, 3rd edition.

\bibitem[\protect\citeauthoryear{Jones and Whitehead}{Jones and Whitehead}{1979}]{jones1979}
Jones, D. and Whitehead, J. (1979).
\newblock Sequential forms of the log rank and modified wilcoxon tests for censored data.
\newblock {\em Biometrika} {\bf 66,} 105--113.

\bibitem[\protect\citeauthoryear{Li and Shepherd}{Li and Shepherd}{2012}]{Li2012}
Li, C. and Shepherd, B.~E. (2012).
\newblock A new residual for ordinal outcomes.
\newblock {\em Biometrika} {\bf 99,} 473--480.

\bibitem[\protect\citeauthoryear{Pearson}{Pearson}{1907}]{Pearson1907}
Pearson, K.~G. (1907).
\newblock {\em On Further Methods of Determining Correlation}.
\newblock Cambridge University Press, Cambridge.

\bibitem[\protect\citeauthoryear{Rosner and Glynn}{Rosner and Glynn}{2011}]{rosner2011}
Rosner, B. and Glynn, R. (2011).
\newblock Power and sample size estimation for the clustered wilcoxon test.
\newblock {\em Biometrics} {\bf 67,} 646–653.

\bibitem[\protect\citeauthoryear{Tu, Li, Zeng, and Shepherd}{Tu et~al.}{2023}]{tu2023}
Tu, S., Li, C., Zeng, D., and Shepherd, B.~E. (2023).
\newblock Rank intraclass correlation for clustered data.
\newblock {\em Statistics in Medicine} {\bf 42,} 4333--4348.

\bibitem[\protect\citeauthoryear{Whitehead}{Whitehead}{1993}]{whitehead1993}
Whitehead, J. (1993).
\newblock Sample size calculations for ordered categorical data.
\newblock {\em Statistics in Medicine} {\bf 12,} 2257--2271.

\end{thebibliography}

\section*{Supplementary Materials}
Web Appendices and Figures referenced in Sections \ref{s:review}, \ref{s:calculation}, and \ref{s:simulations} are available with this paper at the publisher's website.

\clearpage

\clearpage

\clearpage 

\clearpage 

\clearpage 

\clearpage 

\end{document}


\maketitle
\newpage




\label{firstpage}

\maketitle

\section{Web Appendix A: Score statistics and their variances of unadjusted proportional odds models for ordinal and continuous outcomes}
In this appendix, we provide the analytical rationale for our sample size calculation for continuous outcomes, which is an extension of Whitehead's sample size formula for ordinal data by treating the proportion of each category as $1/N$, where $N$ is the total sample size. Whitehead's sample size formula is based on an approximation of the variance of the score statistic derived from an unadjusted proportional odds (PO) model with an ordinal variable as the outcome and the treatment indicator as the predictor. In the following, we will show that the score statistic derived from an unadjusted cumulative probability model (CPM) of a continuous outcome is the same as the score statistic underlying Whitehead's formula. We will also show that Whitehead's variance approximation for the score statistic continues to hold for continuous outcomes if $\bar{\pi}_{l}$ is replaced with $1/N$, and that the approximation is better for continuous data than it is for ordinal data with only a few categories. 

Consider a family of cumulative distribution functions (CDFs) on a latent continuous scale: $F(s;\theta)=\frac{H(s)}{H(s)+(1-H(s))e^{\frac{1}{2}\theta}}$, where $s\in R$, $H(s)$ is an unspecified baseline CDF. If $\frac{1}{2}\theta=\beta X$, this is another way of expressing proportional
odds CPM. [Proof: In CPM, $F_Y(y|X) =P(Y\le y|X) =G^{-1}(\alpha(y)-\beta X)$, where $\alpha(y)$ is an increasing function. With the logit link, $G^{-1}(x)=\frac{e^x}{1+e^x}$, and the odds function of $F_Y(y|X)$ is
\begin{equation*}
  \frac{1-F_Y(y|X)}{F_Y(y|X)}
  =\frac{1-G^{-1}(\alpha(y)-\beta X)}{G^{-1}(\alpha(y)-\beta X)}
  =e^{-(\alpha(y)-\beta X)} =e^{-\alpha(y)}\cdot e^{\beta X}.
\end{equation*}
Let $H(y)=\frac{1}{1+e^{-\alpha(y)}}$. $H(y)$ is the baseline CDF (when $X=0$), and $e^{-\alpha(y)} =\frac{1-H(y)}{H(y)}$ is the corresponding baseline odds function.  Then we have the proportional odds expression
\begin{equation*}
  \frac{1-F_Y(y|X)}{F_Y(y|X)}
  =\frac{1-H(y)}{H(y)}\cdot e^{\beta X} =A.
\end{equation*}
Solving this for $F_Y(y|X)$, we obtain $F_Y(y|X)=\frac{1}{1+A} =\frac{H(y)}{H(y)+(1-H(y))e^{\beta X}}$.]

For two-group continuous data, we assume the CDF is $F(s;\theta)$ for group 1 and $F(s;-\theta)$ for group 2. Let $m$ and $n$ be the sample sizes for groups 1 and 2, respectively, and $N=m+n$.  We assume the $N$ observations are independent and there are no ties. Let $s_k$ be the observation that has rank $k$. Let $u_k=H(s_k)$. Then
$0<u_1<u_2<\cdots<u_N<1$. The rank-based marginal likelihood is
\begin{equation*}
  L(\theta) =P(0<u_1<u_2<\cdots<u_N<1; \theta)
  =\int_R dF_1(u_1)dF_2(u_2)\cdots dF_N(u_N),
\end{equation*}
where $R=\{(u_1,\ldots,u_N): 0<u_1<\cdots<u_N<1\} \subset R^N$; and $F_k(u)$
is the CDF for $u_k$: $F_k(u)=\frac{u}{u+(1-u)e^{\frac{1}{2}\theta}}$ if it is in group 1 and $F_k(u)=\frac{u}{u+(1-u)e^{-\frac{1}{2}\theta}}$ if it is in group 2.  Their corresponding PDFs are $\frac{e^{\frac{1}{2}\theta}}{(u+(1-u)e^{\frac{1}{2}\theta})^2}$ and $\frac{e^{-\frac{1}{2}\theta}}{(u+(1-u)e^{-\frac{1}{2}\theta})^2}$,
respectively.  Thus the likelihood above can be expressed as:
\begin{equation*}
  L(\theta) =\int_R \frac{ e^{\frac{1}{2}(m-n)\theta} }{P(\theta)} du_1du_2\cdots du_N,
\end{equation*}
where
\begin{equation*}
  P(\theta) =\prod_1 \{u_k+(1-u_k)e^{\frac{1}{2}\theta}\}^2 \prod_2
  \{u_k+(1-u_k)e^{-\frac{1}{2}\theta}\}^2,
\end{equation*}
and $\prod_j$ denotes a product over the observations in group $j$. Similar usage for $\sum_j$ below. Since
$\frac{d\log L(\theta)}{d\theta} =\frac{L'(\theta)}{L(\theta)}$ and $\frac{d^2\log L(\theta)}{d\theta^2} =\frac{L''(\theta)}{L(\theta)}
-\frac{L'(\theta)^2}{L(\theta)^2}$, the score and the Fisher information are
\begin{equation*}
  Z =\frac{L'(0)}{L(0)}, \quad
  V=\frac{L'(0)^2}{L(0)^2} -\frac{L''(0)}{L(0)} =Z^2 -\frac{L''(0)}{L(0)}.
\end{equation*}
Note that $L(0) =\int_R du_1\cdots du_N =\frac{1}{N!}$ when there are no ties (derived below).

When there are ties, the region $R\subset R^N$ will be defined accordingly. For example, suppose $N=6$, and there are 3 distinct outcome categories, with 2 observations at the lowest category, 1 the middle category, and 3 the
highest category.  In this situation, $R=\{(u_1,\ldots,u_6): 0<u_1=u_2<u_3<u_4=u_5=u_6<1\}$, and the likelihood should be multiplied by $2!1!3!$.  In general, if there are $A$ categories and $c_a$ observations for category $a$, the likelihood should be multiplied
by $\prod_a(c_a!)$. Ties have no impact on $Z$ and $V$ because the constant multiplier is canceled in the ratios.

We can derive $L'(\theta)$ and $L''(\theta)$ if integral and derivative are exchangeable. They are
\begin{align*}
  L'(\theta)
  &=\int_R\left\{
    \frac{e^{\frac{1}{2}(m-n)\theta}\frac{1}{2}(m-n)}{P(\theta)}
    -\frac{e^{\frac{1}{2}(m-n)\theta}P'(\theta)}{P(\theta)^2}
    \right\} du_1\cdots du_N, \\
  L''(\theta)
  &=\int_R\left[ \frac{1}{2}(m-n) \left\{
    \frac{e^{\frac{1}{2}(m-n)\theta}\frac{1}{2}(m-n)}{P(\theta)}
    -\frac{e^{\frac{1}{2}(m-n)\theta}P'(\theta)}{P(\theta)^2}
    \right\} \right. \\
  &\qquad\quad -\frac{e^{\frac{1}{2}(m-n)\theta}P'(\theta)}{P(\theta)^2}
    \left. \left\{
    \frac{1}{2}(m-n)
    +\frac{P''(\theta)}{P'(\theta)}
    -\frac{2P(\theta)P'(\theta)}{P(\theta)^2}
    \right\} \right]
    du_1\cdots du_N.
\end{align*}
Since $P(0)=1$, we have
\begin{align*}
  L'(0)
  &=\int_R\left\{
    \frac{1}{2}(m-n) -P'(0)
    \right\} du_1\cdots du_N, \\
  L''(0)
  &=\int_R\left[
    \frac{1}{2}(m-n)
    \left\{
    \frac{1}{2}(m-n) -P'(0)
    \right\} \right. \\
  &\qquad\quad -P'(0) \left. \left\{
    \frac{1}{2}(m-n)
    +\frac{P''(0)}{P'(0)}
    -2P'(0)
    \right\} \right] du_1\cdots du_N \\
  &=\int_R\left\{
    \frac{1}{4}(m-n)^2
    -(m-n)P'(0)
    -P''(0)
    +2P'(0)^2
    \right\} du_1\cdots du_N.
\end{align*}
We now derive $P'(\theta)$ and $P''(\theta)$.  $P'(\theta)$ is
\begin{equation*}
  P'(\theta) =P(\theta) \left[
    \sum_1\frac{(1-u_k)e^{\frac{1}{2}\theta}}{u_k+(1-u_k)e^{\frac{1}{2}\theta}}
    -\sum_2\frac{(1-u_k)e^{-\frac{1}{2}\theta}}{u_k+(1-u_k)e^{-\frac{1}{2}\theta}}
  \right].
\end{equation*}
To derive $P''(\theta)$, we note that
$\frac{(1-u_k)e^{\frac{1}{2}\theta}}{u_k+(1-u_k)e^{\frac{1}{2}\theta}}
=1-\frac{u_k}{u_k+(1-u_k)e^{\frac{1}{2}\theta}}$ and thus its derivative with respect to $\theta$ is
$\frac{u_k(1-u_k)e^{\frac{1}{2}\theta}\frac{1}{2}}{(u_k+(1-u_k)e^{\frac{1}{2}\theta})^2}$; and similarly,
$\frac{(1-u_k)e^{-\frac{1}{2}\theta}}{u_k+(1-u_k)e^{-\frac{1}{2}\theta}}
=1-\frac{u_k}{u_k+(1-u_k)e^{-\frac{1}{2}\theta}}$ and thus its derivative with respect to $\theta$ is
$-\frac{u_k(1-u_k)e^{-\frac{1}{2}\theta}\frac{1}{2}}{(u_k+(1-u_k)e^{-\frac{1}{2}\theta})^2}$. 
Then,
\begin{align*}
  P''(\theta)
  &=P'(\theta) \left[
    \sum_1\frac{(1-u_k)e^{\frac{1}{2}\theta}}{u_k+(1-u_k)e^{\frac{1}{2}\theta}}
    -\sum_2\frac{(1-u_k)e^{-\frac{1}{2}\theta}}{u_k+(1-u_k)e^{-\frac{1}{2}\theta}}
    \right] \\
  &\quad +P(\theta) \left[
    \sum_1
    \frac{u_k(1-u_k)e^{\frac{1}{2}\theta}\frac{1}{2}}{(u_k+(1-u_k)e^{\frac{1}{2}\theta})^2}
    +\sum_2
    \frac{u_k(1-u_k)e^{-\frac{1}{2}\theta}\frac{1}{2}}{(u_k+(1-u_k)e^{-\frac{1}{2}\theta})^2}
    \right].
\end{align*}
Thus,
\begin{align*}
  P'(0)
  &=\sum_1(1-u_k)-\sum_2(1-u_k) =(m-n) -(\sum_1u_k-\sum_2u_k), \\
  P''(0)
  &=P'(0)^2
    +\frac{1}{2}\sum u_k(1-u_k).
\end{align*}

We now work out the pieces of integrals in $L'(0)$ and $L''(0)$. It can be shown that
\begin{align*}
  \int_R du_1\cdots du_N &=\frac{1}{N!}, \\
  \int_R u_k\ du_1\cdots du_N &=\frac{k}{(N+1)!}, \\
  \int_R u_k^2\ du_1\cdots du_N &=\frac{k(k+1)}{(N+2)!}, \\
  \int_R u_ju_k\ du_1\cdots du_N &=\frac{j(k+1)}{(N+2)!}\quad (\text{when } j<k).
\end{align*}
For example, the proof of the first equation is
\begin{align*}
  \int_R du_1\cdots du_N
  &=\int_0^1\int_0^{u_N}\cdots\int_0^{u_3}\int_0^{u_2}
    du_1du_2\cdots du_{N-1}du_N \\
  &=\int_0^1\int_0^{u_N}\cdots\int_0^{u_3} u_2\ du_2\cdots du_{N-1}du_N \\
  &=\int_0^1\int_0^{u_N}\cdots\int_0^{u_4} \frac{1}{2}u_3^2\ du_3\cdots
    du_{N-1}du_N \\
  &=\int_0^1\int_0^{u_N}\cdots\int_0^{u_5} \frac{1}{3!}u_4^3\ du_4\cdots
    du_{N-1}du_N \\
  &=\cdots =\int_0^1 \frac{1}{(N-1)!}u_N^{N-1}\ du_N \\
  &=\frac{1}{N!}.
\end{align*}
Then
\begin{align*}
  \int_R \sum u_k(1-u_k)\ du_1\cdots du_N 
  &=\sum \left\{ \frac{k}{(N+1)!} -\frac{k(k+1)}{(N+2)!} \right\} \\
  &=\frac{1}{(N+2)!}\sum k(N+1-k) =\frac{1}{6(N-1)!}; \\
  \int_R P'(0)\ du_1\cdots du_N 
  &=\int_R\left\{
    (m-n)-(\sum_1u_k-\sum_2u_k)
    \right\} du_1\cdots du_N \\
  &=(m-n)\frac{1}{N!} -\frac{1}{(N+1)!}(\sum_1k-\sum_2k); \\
  \int_R P'(0)^2\ du_1\cdots du_N 
  &=\int_R\left\{
    (m-n)-(\sum_1u_k-\sum_2u_k)
    \right\}^2 du_1\cdots du_N \\
  &=(m-n)^2\frac{1}{N!} -\frac{2(m-n)}{(N+1)!}(\sum_1k-\sum_2k) \\
  &\quad +\frac{1}{(N+2)!}(\sum k(k+1)+2\sum_{j<k}j(k+1)\lambda_{j,k}),
\end{align*}
where $\lambda_{j,k}=1$ if $j$ and $k$ are in the same group and
$\lambda_{j,k}=-1$ otherwise. Then we have:
\begin{align*}
  L'(0)
  &=\int_R\left\{
    \frac{1}{2}(m-n) -P'(0)
    \right\} du_1\cdots du_N \\
  &=\frac{1}{(N+1)!}(\sum_1k-\sum_2k) 
    -\frac{1}{2}(m-n)\frac{1}{N!} \\
  &=\frac{1}{(N+1)!} \left\{ \sum_1k-\sum_2k -\frac{1}{2}(m-n)(N+1) \right\}.
\end{align*}

Let $A$ be the number of distinct outcome categories.  For $a=1,\ldots,A$, let $m_a$ and $n_a$ be the numbers of observations with category $a$ in groups 1 and 2, respectively. Then $m=\sum_am_a$ and $n=\sum_an_a$. Let
$c_a=m_a+n_a$. Let $L_a$ be the cumulative count from the lower end up to, but not including, category $a$; that is, $L_1=0$, and $L_a=c_1+\cdots+c_{a-1}$ for $a=2,\ldots,A$.  Similarly, let $U_a$ be the cumulative count from the upper end up to, but not including, category $a$; that is, $U_A=0$, and $U_a=c_{a+1}+\cdots+c_A$ for $a=1,\ldots,A-1$. Then $L_a+c_a+U_a=N$ for any $a$.  After the data are ranked, the midrank for category $a$ is $r_a=L_a+\frac{1+c_a}{2}$. Then $\sum_1 k=\sum_a m_ar_a$,
and $\sum_1 k+\sum_2 k =\sum k =\frac{1}{2}N(N+1)$. The score is
\begin{align*}
  Z =\frac{L'(0)}{L(0)}
  &=\frac{1}{N+1} \left\{ \sum_1k-\sum_2k -\frac{1}{2}(m-n)(N+1) \right\} \\
  &=\frac{1}{N+1} \left\{ 2\sum_1k-\frac{1}{2}N(N+1) -\frac{1}{2}(m-n)(N+1) \right\} \\
  &=\frac{1}{N+1} \left\{ \sum_a 2m_ar_a-m(N+1) \right\} \\
  &=\frac{1}{N+1} \sum_a m_a(2r_a-(N+1)) \\
  &=\frac{1}{N+1} \sum_a m_a(2L_a+c_a-N) \\
  &=\frac{1}{N+1} \sum_a m_a(L_a-U_a),
\end{align*}
which is equivalent to the score statistic underlying Whitehead's sample size formula \cite{whitehead1993}. 

We also have:
\begin{align*}
  L''(0)
  &=\int_R\left\{
    \frac{1}{4}(m-n)^2
    -(m-n)P'(0)
    -P''(0)
    +2P'(0)^2
    \right\} du_1\cdots du_N \\
  &=\int_R\left\{
    \frac{1}{4}(m-n)^2
    -(m-n)P'(0)
    +P'(0)^2
    -\frac{1}{2}\sum u_k(1-u_k)
    \right\} du_1\cdots du_N \\
  &=\frac{1}{4}(m-n)^2\frac{1}{N!} -(m-n)\left\{ (m-n)\frac{1}{N!}
    -\frac{1}{(N+1)!}(\sum_1k-\sum_2k) \right\} \\
  &\quad +\left\{ (m-n)^2\frac{1}{N!} -\frac{2(m-n)}{(N+1)!}(\sum_1k-\sum_2k)
    +\frac{1}{(N+2)!}(\sum k(k+1)+2\sum_{j<k}j(k+1)\lambda_{j,k}) \right\} \\
  &\quad -\frac{1}{12}\frac{1}{(N-1)!} \\
  &=\frac{(m-n)^2}{4N!} -\frac{(m-n)}{(N+1)!}(\sum_1k-\sum_2k)
    +\frac{1}{4(N-1)!}
    +\frac{2}{(N+2)!} \sum_{j<k}j(k+1)\lambda_{j,k}.
\end{align*}
During the last step, we used the fact that
$\sum k(k+1) =\frac{1}{3}N(N+1)(N+2)$. With the Appendix of Jones and Whitehead's paper \cite{jones1979}, we can have
\begin{align*}
  V &=Z^2-\frac{L''(0)}{L(0)} \\
  &=\frac{1}{(N+1)^2} (\sum_{i,j} U_{ij})^2
    -\frac{1}{(N+1)(N+2)} (\sum_i\sum_{j\ne k} V_{ijk}
    +\sum_{j}\sum_{i\ne h} W_{ijh} +\sum_{i\ne h}\sum_{j\ne k} U_{ij}U_{hk}) \\
  &=\frac{1}{(N+1)^2} \sum_{i,j} U_{ij}^2
    +\frac{1}{(N+1)^2} (\sum_i\sum_{j\ne k} U_{ij}U_{ik}
    +\sum_{j}\sum_{i\ne h} U_{ij}U_{hj} +\sum_{i\ne h}\sum_{j\ne k} U_{ij}U_{hk}) \\
  &\quad -\frac{1}{(N+1)(N+2)} (\sum_i\sum_{j\ne k} V_{ijk}
    +\sum_{j}\sum_{i\ne h} W_{ijh} +\sum_{i\ne h}\sum_{j\ne k} U_{ij}U_{hk}) \\
  &= T_1+T_2+T_3+T_4+T_5+T_6+T_7,
\end{align*}
where the seven terms are
\begin{align*}
  &T_1= \frac{1}{(N+1)^2} \sum_{i,j} U_{ij}^2
  =\frac{1}{(N+1)^2} \sum_i n_{iC}(n_E-n_{iE}), \\
  &T_2= \frac{1}{(N+1)^2} \sum_i\sum_{j\ne k} U_{ij}U_{ik}
  = \frac{1}{(N+1)^2} \sum_i n_{iC}(L_{iE}-U_{iE})^2, \\
  &T_3= \frac{1}{(N+1)^2} \sum_{j}\sum_{i\ne h} U_{ij}U_{hj}
  = \frac{1}{(N+1)^2} \sum_i n_{iE}(L_{iC}-U_{iC})^2, \\
  &T_5  = \frac{-1}{(N+1)(N+2)} \sum_i\sum_{j\ne k} V_{ijk}\\
  &= \frac{-1}{(N+1)(N+2)} \sum_{i} n_{iC}(L_{iE}^2+U_{iE}^2
    -2n_{iE}(n_E-n_{iE}) -4L_{iE}U_{iE}), \\
  &T_6  = \frac{-1}{(N+1)(N+2)} \sum_{j}\sum_{i\ne h} W_{ijh} \\
  &= \frac{-1}{(N+1)(N+2)} \sum_{i} n_{iE}(L_{iC}^2+U_{iC}^2
    -2n_{iC}(n_C-n_{iC}) -4L_{iC}U_{iC}),
\end{align*}
and
\begin{align*}
  T_4+T_7 &= \frac{1}{(N+1)^2(N+2)} \sum_{i\ne h}\sum_{j\ne k} U_{ij}U_{hk}\\
  &= \frac{1}{(N+1)^2(N+2)} \sum_{i\ne h}
    n_{iC}n_{hC}(L_{iE}-U_{iE})(L_{hE}-U_{hE})\\ &=T_8.
\end{align*}
Since
\begin{align*}
  T_2+T_5
  &= \frac{1}{(N+1)^2} \sum_i n_{iC}(L_{iE}-U_{iE})^2 \\
  & \;\;\;\;  -\frac{1}{(N+1)(N+2)} \sum_{i} n_{iC}(L_{iE}^2+U_{iE}^2
    -2n_{iE}(n_E-n_{iE}) -4L_{iE}U_{iE}), \\
  &= \frac{1}{(N+1)^2(N+2)} \sum_i n_{iC}(L_{iE}-U_{iE})^2 \\
   & \;\;\;\; +\frac{2}{(N+1)(N+2)} \sum_{i} n_{iC}(n_{iE}(n_E-n_{iE})+L_{iE}U_{iE}),
  \\
  &= T_9+T_{10},
\end{align*}
and similarly,
\begin{align*}
  T_3+T_6
  &= \frac{1}{(N+1)^2(N+2)} \sum_i n_{iE}(L_{iC}-U_{iC})^2\\
  & \;\;\;\;  +\frac{2}{(N+1)(N+2)} \sum_{i} n_{iE}(n_{iC}(n_C-n_{iC})+L_{iC}U_{iC}),
  \\
  &= T_{11}+T_{12},
\end{align*}  
we have
$$V=T_1+T_9+T_{10}+T_{11}+T_{12}+T_8.$$  
Under the null and with $N$ large, $n_{iC}\approx n_Cp_i$,
$n_{iE}\approx n_Ep_i$, $L_{iE}\approx n_E\gamma_{i-1}$,
$U_{iE}\approx n_E(1-\gamma_i)$, etc., where $p_i=c_i/N$ and
$\gamma_i=p_1+\cdots+p_i$. When $N$ is large, $T_{10},T_{12},T_8\propto N$, but $T_1,T_9,T_{11}$ are bounded by $1$ and can be ignored (e.g., $T_1\approx \frac{n_Cn_E}{(N+1)^2} \sum_i p_i(1-p_i) =\frac{n_Cn_E}{(N+1)^2}
(1-\sum_i p_i^2) <1$). Thus,
\begin{align*}
  V
  &\approx T_{10}+T_{12}+T_8 \\
  &\approx \frac{2n_Cn_E^2}{(N+1)(N+2)}
    \sum_{i} p_i(p_i(1-p_i)+\gamma_{i-1}(1-\gamma_i)) \\
  &\quad +\frac{2n_C^2n_E}{(N+1)(N+2)}
    \sum_{i} p_i(p_i(1-p_i)+\gamma_{i-1}(1-\gamma_i)) \\
  &\quad +\frac{n_C^2n_E^2}{(N+1)^2(N+2)} \sum_{i\ne h}
    p_ip_h(\gamma_{i-1}-(1-\gamma_i))(\gamma_{h-1}-(1-\gamma_h)) \\
  &= \frac{n_Cn_EN}{(N+1)(N+2)}
    \sum_{i} (2p_i^2(1-p_i)+2p_i\gamma_{i-1}(1-\gamma_i)) \\
  &\quad -\frac{n_C^2n_E^2}{(N+1)^2(N+2)} \sum_i
    p_i^2(\gamma_{i-1}-(1-\gamma_i))^2 \
    \quad (\text{because } \sum_{h} p_h(\gamma_{h-1}-(1-\gamma_h))=0) \\
  &= \frac{n_Cn_EN}{(N+1)(N+2)}
    \sum_{i} (p_i^2(1-p_i)+2p_i\gamma_{i-1}(1-\gamma_i)) \\
  &\quad +\frac{n_Cn_EN}{(N+1)(N+2)}
    \sum_{i} p_i^2(1-p_i) -\frac{n_C^2n_E^2}{(N+1)^2(N+2)} \sum_i
    p_i^2(\gamma_{i-1}-(1-\gamma_i))^2 \\
  &= \frac{n_Cn_EN}{(N+1)^2}
    \sum_{i} (p_i^2(1-p_i)+2p_i\gamma_{i-1}(1-\gamma_i)) \\
  &\quad -\frac{n_Cn_EN}{(N+1)^2(N+2)}
    \sum_{i} (p_i^2(1-p_i)+2p_i\gamma_{i-1}(1-\gamma_i)) \\
  &\quad +\frac{n_Cn_EN}{(N+1)(N+2)}
    \sum_{i} p_i^2(1-p_i) -\frac{n_C^2n_E^2}{(N+1)^2(N+2)} \sum_i
    p_i^2(\gamma_{i-1}-(1-\gamma_i))^2.
\end{align*}
Let $V_1$ be the first term in the last equation.  Since
$\sum_i p_i^2(1-p_i) +2\sum_i p_i\gamma_{i-1}(1-\gamma_i) =\sum_{a\ne
  b}p_ap_b^2 +2\sum_{a<b<c}p_ap_bp_c =(1-\sum p_i^3)/3$ (details in \cite{Li2012}), $V_1=\frac{n_Cn_EN}{3(N+1)^2}\left( 1-\sum_i p_i^3 \right)$,
the equation (3) in \cite{whitehead1993}. The remaining terms become
\begin{align*}
  &\frac{n_Cn_E}{(N+1)^2(N+2)}
    \left[ N(N+1)\sum_{i} p_i^2(1-p_i) -n_Cn_E \sum_i
    p_i^2(\gamma_{i-1}-(1-\gamma_i))^2 \right. \\
  &\qquad\qquad\qquad\qquad
    \left. -N\sum_{i} (p_i^2(1-p_i)+2p_i\gamma_{i-1}(1-\gamma_i))  \right] \\
  =\ &\frac{n_Cn_E}{(N+1)^2(N+2)}
       \left[ N^2\sum_{i} p_i^2(1-p_i) -n_Cn_E \sum_i
       p_i^2(\gamma_{i-1}-(1-\gamma_i))^2 \right. \\
  &\qquad\qquad\qquad\qquad
    \left. -N\sum_{i} (2p_i\gamma_{i-1}(1-\gamma_i))  \right] \\
  &=V_2+V_3+V_4.
\end{align*}
Here $V_4$ is bounded and can be ignored, but the other two terms are $\propto N$.

We now evaluate the contributions of $V_2$ and $V_3$ relative to that of $V_1$. Let $A$ be the number of categories and suppose $p_i\equiv 1/A$.
Then $V_1 =\frac{n_Cn_E}{(N+1)^2}\cdot\frac{N}{3}(1-\frac{1}{A^2})$,
$V_2=\frac{n_Cn_E}{(N+1)^2}\cdot\frac{N^2}{N+2}(\frac{1}{A}-\frac{1}{A^2})$,
and
$V_3=-\frac{n_Cn_E}{(N+1)^2}\cdot\frac{n_Cn_E}{N+2}\frac{1}{3A}(1-\frac{1}{A^2})$.
When $A=3$, the three factors are $\frac{24N}{81}$, $\frac{18N}{81}$, $-\frac{8n_Cn_E}{81N}=-\frac{2N}{81}$ if $n_C=n_E=N/2$.  When $A=5$, ignoring the $\frac{1}{A^2}$ and $\frac{1}{A^3}$ parts, the three factors are approximately $\frac{N}{3}$, $\frac{N}{5}$, $-\frac{n_Cn_E}{15N}$.  When
$A=10$, they are approximately $\frac{N}{3}$, $\frac{N}{10}$,
$-\frac{n_Cn_E}{30N}$. So, when $A$ is large, the Whitehead approximation works well, but when $A$ is very small (e.g., $A\le5$), the Whitehead approximation ignores significant contributions from $V_2$ and $V_3$ (e.g.,
when $A=3$, $V_1\propto\frac{24N}{81}$, while
$V_2+V_3\propto\frac{16N}{81}$).

In sample size calculations, the formula for $V$ as a function of $N$ is set to equal $[\frac{u_{1-\alpha/2}+u_\beta}{\theta_R}]^2$ (which is $var(Z)$ needed given the values of $\alpha$, $\beta$, and $\theta_R$) to derive the desired $N$.  When $A$ is small, because the Whitehead formula ignores the contribution from $V_2+V_3$, it has to have a larger $N$ to meet the required $var(Z)$. This likely explains the inflated power in the simulations for ordinal outcomes with a very small number of categories. However, for a large number of categories (e.g., when the outcome is continuous), $V_1$, which is the formula in Whitehead, serves as a good approximation for the variance $V$.


\section{Web Appendix B: The design effect of cluster RCTs associated with the clustered Wilcoxon rank-sum test statistic}
\label{s:de} 
With continuous outcomes, Rosner and Glynn \citep{rosner2011} derived $\text{var}(\hat{\theta}_{\text{\scriptsize IR}})$ for independent data and $\text{var}(\hat \theta)$ for clustered data. Then we can derive the DE of cluster RCTs associated with $\hat \theta$ for continuous outcomes. Let $\rho$ denote the ICC after the probit transformation on the cumulative distribution function. Let $Q(\theta, \rho)=\Phi_2(\Phi^{-1}(\theta),\Phi^{-1}(\theta),\rho) -\theta^2$, where $\Phi_2(\Phi^{-1}(\theta),\Phi^{-1}(\theta),\rho) = P(Z_1 \leq \Phi^{-1}(\theta), Z_2 \leq \Phi^{-1}(\theta))$ with $(Z_1,Z_2) \sim N(\begin{pmatrix} 0 \\ 0 \end{pmatrix}, \begin{pmatrix} 1 & \rho \\ \rho &  1 \end{pmatrix})$. Let $m_0$ and $m_1$ be the numbers of clusters in the control and experiment arms, and $k$ denote the cluster size. The DE of cluster RCTs associated with $\hat \theta$ for continuous outcomes is 
\begin{equation*}
    \begin{aligned}
    D_{\text{eff}}(\hat \theta) & = \frac{\text{var}(\hat \theta)}{\text{var}(\hat{\theta}_{\text{\scriptsize IR}})} \\
     & = \frac{\Big\{\theta(1-\theta)+2(k-1)Q(\theta,\frac{1+\rho}{2})+(k-1)^2Q(\theta,\rho) \Big\} /m_0m_1k^2 }{\Big\{\theta(1-\theta) + (m_0k+m_1k-2)Q(\theta,\frac{1}{2})\Big\}/m_0m_1k^2 } \\
    & + \frac{\Big\{k(m_0+m_1-2)[Q(\theta,\frac{1}{2})-(k-1)Q(\theta,\frac{\rho}{2})]\Big\}/m_0m_1k^2}{\Big\{\theta(1-\theta) + (m_0k+m_1k-2)Q(\theta,\frac{1}{2})\Big\}/m_0m_1k^2}  \\
    & = \frac{\theta(1-\theta)}{\theta(1-\theta) + (m_0k+m_1k-2)Q(\theta,\frac{1}{2})} + \frac{(2k-2)Q(\theta,\frac{1+\rho}{2})}{\theta(1-\theta) + (m_0k+m_1k-2)Q(\theta,\frac{1}{2})} \\
    & + \frac{(m_0k+m_1k-2k)Q(\theta,\frac{1}{2})}{\theta(1-\theta) + (m_0k+m_1k-2)Q(\theta,\frac{1}{2})} + \frac{(k-1)(m_0k+m_1k-2k)Q(\theta,\frac{\rho}{2})}{\theta(1-\theta) + (m_0k+m_1k-2)Q(\theta,\frac{1}{2})}\\
    & = (a) + (b) + (c) + (d). 
    \end{aligned}
\end{equation*}
Given $\theta \in [0,1]$, we know that $\theta(1-\theta) \leq \frac{1}{2}$. When $m_0, m_1 >> k$, $(a) \approx 0$, $(b)  \approx 0$, $(c)  \approx 1$, and $(d) \approx (k-1)(\theta,\frac{\rho}{2})/Q(\theta,\frac{1}{2})$. Hence, with $m_0, m_1 >> k$,  $D_{\text{eff}}(\hat \theta) \approx 1 + (k-1)Q(\theta,\rho/2)/Q(\theta,1/2)$. When $\theta=1/2$, which is the probabilistic index under the null, $Q(\theta,\rho/2)/Q(\theta,1/2) = 6\text{sin}^{-1}(\rho/2)/\pi$. After the probit transformation, since the distribution is normal, we have $6\text{sin}^{-1}(\rho/2)/\pi=\gamma_I$ \citep{Pearson1907}, where $\gamma_I$ is the rank ICC \citep{tu2023}. Hence, under the null, if $m_0, m_1 >> k$, $D_{\text{eff}}(\hat \theta) \approx 1 + (k-1)\gamma_I$.  

\section{Web Appendix C: Comparison between the conventional sample size calculations and our sample size calculations under normality}
In this section, we analytically compare the conventional approach using the DE based on the ICC with our method under normality. Let $X$ and $Y$ denote random variables from hierarchical distributions of the control and experiment arms, respectively, and $X \sim N(\mu_0, \sigma^2_0)$ and $Y \sim N(\mu_1, \sigma^2_1)$.
Let $n$ denote the total number of individuals for a cluster RCT, and $n_1 = n/(A+1)$ and $n_0 = An/(A+1)$ denote the number of individuals in the experiment and control arms, respectively, where $A$ is the allocation ratio. The conventional approach based on t-tests and the ICC calculates the total sample size for cluster RCTs with a two-sided significance level at $\alpha$ and power at $1-\beta$ as
$$n^* = \frac{(Z_{1-\alpha/2}+Z_{1-\beta})^2(A\sigma^2_1 + \sigma^2_0)(A+1)}{A(\mu_1-\mu_0)^2}\{1+\rho_I(k-1)\},$$
where $\rho_I$ is the ICC and $k$ is the cluster size. There is a relationship between $\theta$ and $\mu_1-\mu_0$ under normality. Because $X - Y \sim N(\mu_0-\mu_1, \sigma^2_0+\sigma^2_1)$, then we have 
\begin{align*}
\theta &= P(X < Y) \\
& = P(X - Y < 0) \\
& = P(\frac{X - Y - (\mu_0-\mu_1)}{\sqrt{\sigma^2_0+\sigma^2_1}} < \frac{- (\mu_0-\mu_1)}{\sqrt{\sigma^2_0+\sigma^2_1}}) \\
& = \Phi(\frac{- (\mu_0-\mu_1)}{\sqrt{\sigma^2_0+\sigma^2_1}}).
\end{align*}
Then $\mu_1-\mu_0 = \Phi^{-1}(\theta)\sqrt{\sigma^2_0+\sigma^2_1}$. As discussed in the main text, $\theta =  exp(\delta)[exp(\delta)-\delta-1]/(exp(\delta)-1)^2$ where $\delta$ is the log-odds ratio \citep{de_neve2019}. For simplicity, we denote $\theta = h(\delta)$. Then we have $\mu_1-\mu_0 = \Phi^{-1}[h(\delta)]\sqrt{\sigma^2_0+\sigma^2_1}$, and 
$$n^*= \frac{(Z_{1-\alpha/2}+Z_{1-\beta})^2(A\sigma^2_1 + \sigma^2_0)(A+1)}{A\{\Phi^{-1}[h(\delta)]\}^2(\sigma^2_1+\sigma^2_0)}\{1+\rho_I(k-1)\}.$$ 
As described in the main text, with continuous outcomes, our method is $$n = \frac{3(A+1)^2(Z_{1-\alpha/2}+Z_{1-\beta})^2/\delta^2}{A(1 -1/n^2)}\{1+\gamma_I(k-1)\}.$$ 
Then we have $n(1-1/n^2)=3(A+1)^2(Z_{1-\alpha/2}+Z_{1-\beta})^2/(A\delta^2)\{1+\gamma_I(k-1)\}$. We compare the conventional approach with our method,
\begin{align*}
\frac{n-(1/n)}{n^*} & = \frac{3(A+1)^2(Z_{1-\alpha/2}+Z_{1-\beta})^2/(A\delta^2)\{1+\gamma_I(k-1)\}}{\frac{(Z_{1-\alpha/2}+Z_{1-\beta})^2(A\sigma^2_1 + \sigma^2_0)(A+1)}{A\{\Phi^{-1}[h(\delta)]\}^2(\sigma^2_1+\sigma^2_0)}\{1+\rho_I(k-1)\}}\\
& = \frac{3(A+1)\{1+\gamma_I(k-1)\}/\delta^2}{\frac{A\sigma^2_1+\sigma^2_0}{\sigma^2_1+\sigma^2_0}\{1+\rho_I(k-1)\}/\{\Phi^{-1}[h(\delta)]\}^2}.
\end{align*}
If A=1 (or $\sigma_1=\sigma_0$), then $\frac{n-(1/n)}{n^*} =  \frac{6\{1+\gamma_I(k-1)\}/\delta^2}{\{1+\rho_I(k-1)\}/\{\Phi^{-1}[h(\delta)]\}^2}$. We show via simulations that $\frac{6\{1+\gamma_I(k-1)\}/\delta^2}{\{1+\rho_I(k-1)\}/\{\Phi^{-1}[h(\delta)]\}^2} \approx 1$, except for very large $\delta$. Hence, when $A=1$ and $\delta$ is not very large, we can have $n \approx n^*$ since $1/n \approx 0$.

\newpage
\beginsupplement
\section{Web Figure 1}
\begin{figure}[!h]
     \centering
     \begin{subfigure}[b]{0.7\textwidth}
         \centering 
         \includegraphics[width=1\textwidth]{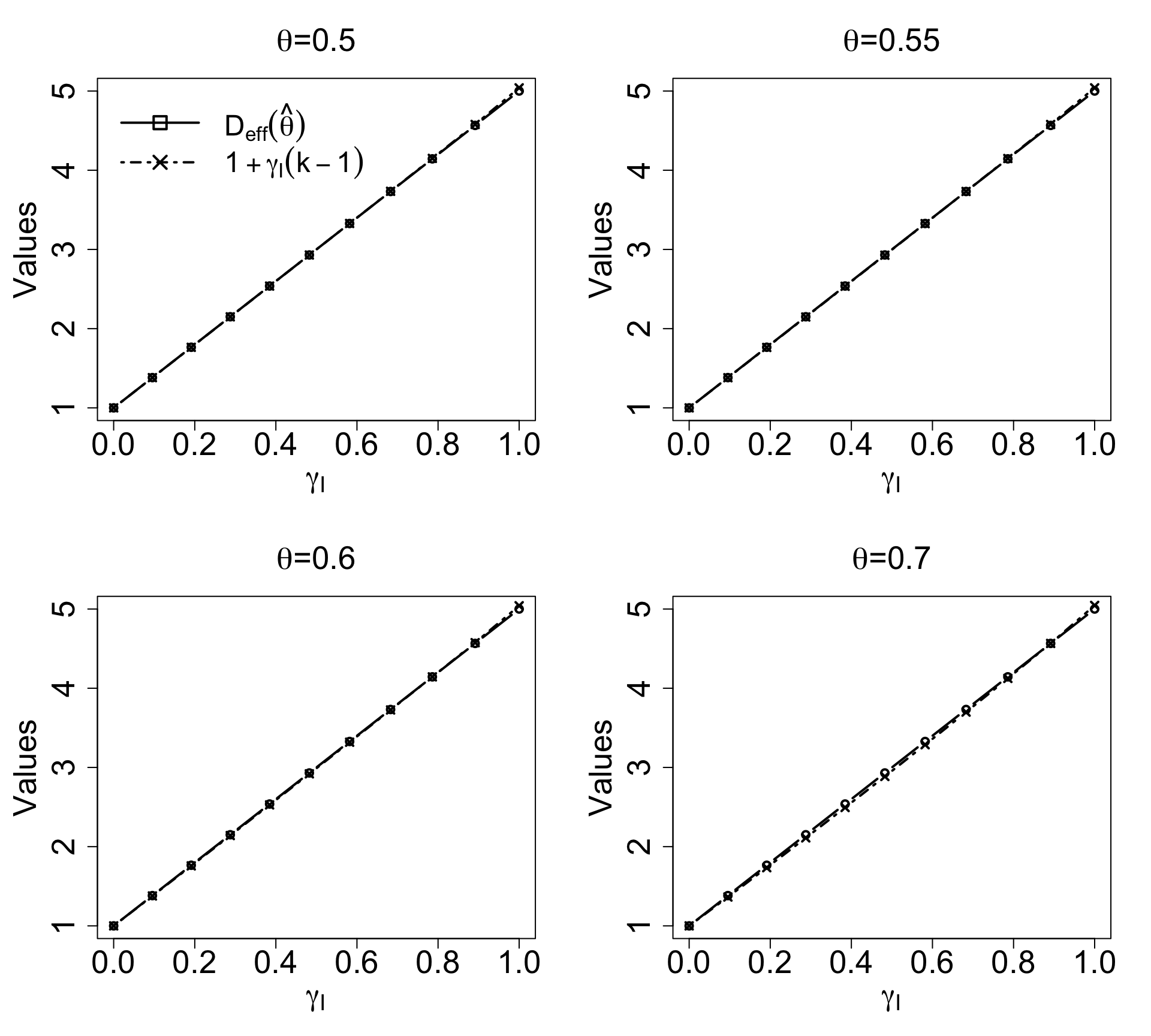} \vspace{-0.6cm}
         \caption{50 clusters and 5 per cluster} 
         \label{fig:DEn50k5} \vspace{0.1cm}
     \end{subfigure}\\
          \begin{subfigure}[b]{0.7\textwidth}
         \centering
         \includegraphics[width=1\textwidth]{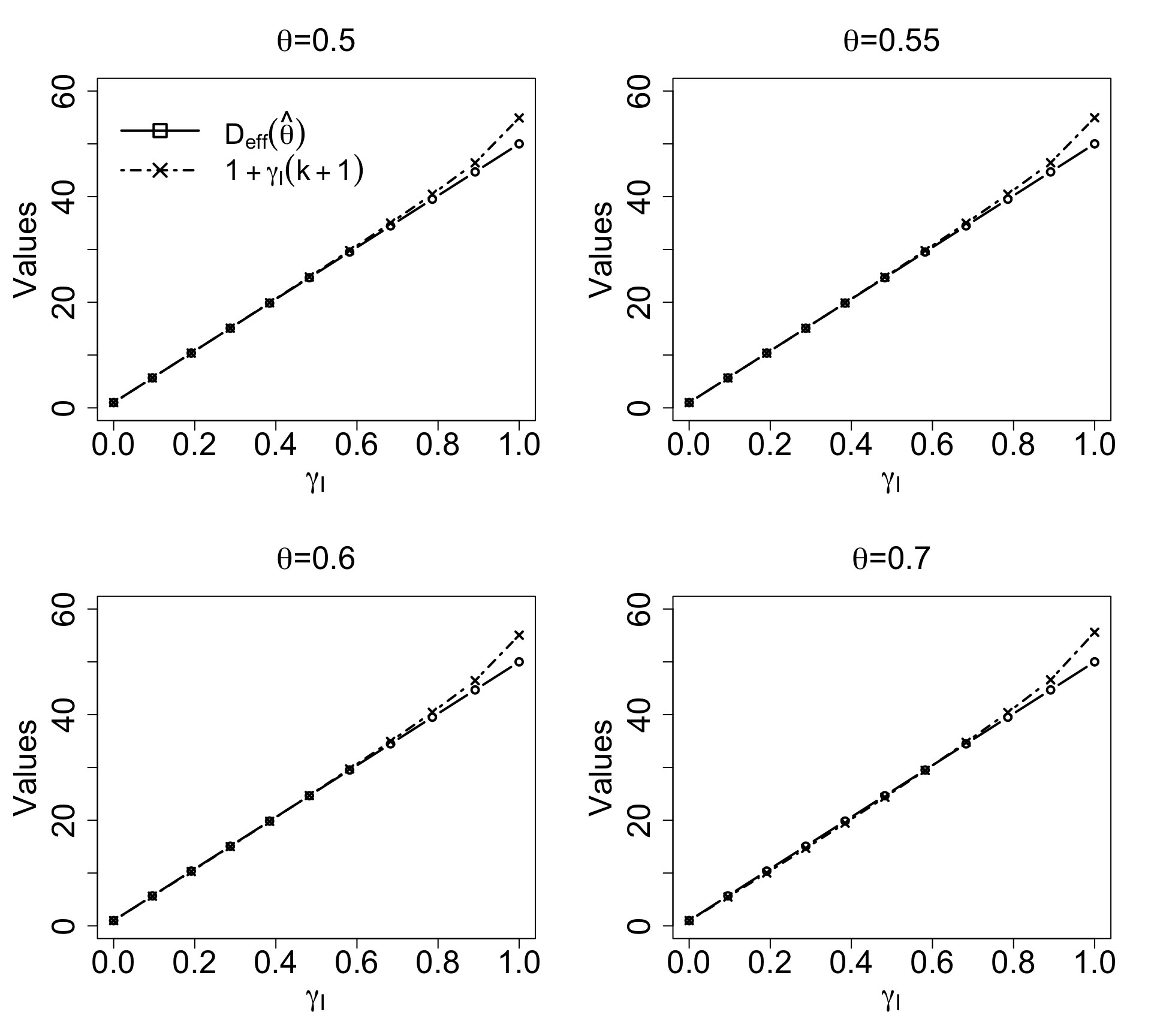} \vspace{-0.6cm}
         \caption{5 clusters and 50 per cluster}
         \label{fig:DEn5k50} \vspace{0.1cm}
     \end{subfigure}
     \vspace{-0.3cm}
     \caption[$D_{\text{eff}}(\hat{\theta})$ versus $1+\gamma_I(k-1)$]{The values of $D_{\text{eff}}(\hat{\theta})$ and $1+\gamma_I(k-1)$ over different values of $\gamma_I$ and $\theta$.}
     \label{fig:DE}
\end{figure}

 We conducted simulations to compare $D_{\text{eff}}(\hat{\theta})$ and $1+\gamma_I(k-1)$ under different values of $\theta$, considering scenarios where the number of clusters was smaller than the cluster size (Figure \ref{fig:DE}). When $\theta=1/2$, $ D_{\text{eff}}(\hat{\theta})$ is close to $1+\gamma_I(k-1)$ even when the cluster size is larger than the number of clusters. When $\theta > 1/2$ and the number of clusters is greater than the cluster size, they are also close, except for large $\gamma_I$.   

\clearpage
\section{Web Figure 2}
\begin{figure}[!h]
     \centering
    \includegraphics[width=1\textwidth]{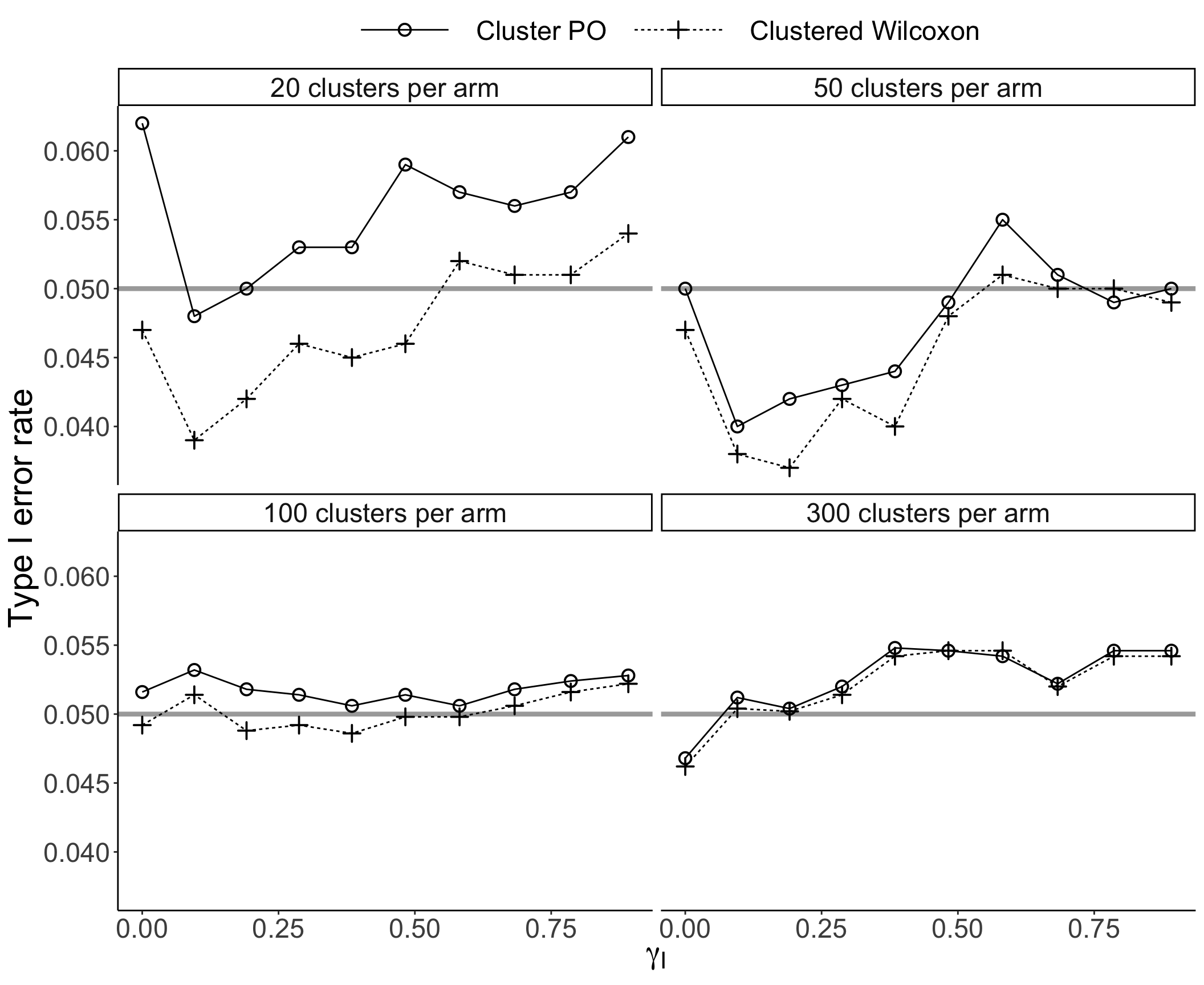} 
     \caption[Type I error of clustered Wilcoxon rank-sum tests and unadjusted cluster PO models across various values of $\gamma_I$]{Type I error rates of clustered Wilcoxon rank-sum tests and unadjusted cluster PO models with the rank ICC $\gamma_I$ varying over $[0, 0.9]$. The cluster size was set to 5.}
     \label{fig:typeI}
\end{figure}

\clearpage 
\section{Web Figure 3} 
\begin{figure}[!h]
    \centering
    \includegraphics[width=1\textwidth]{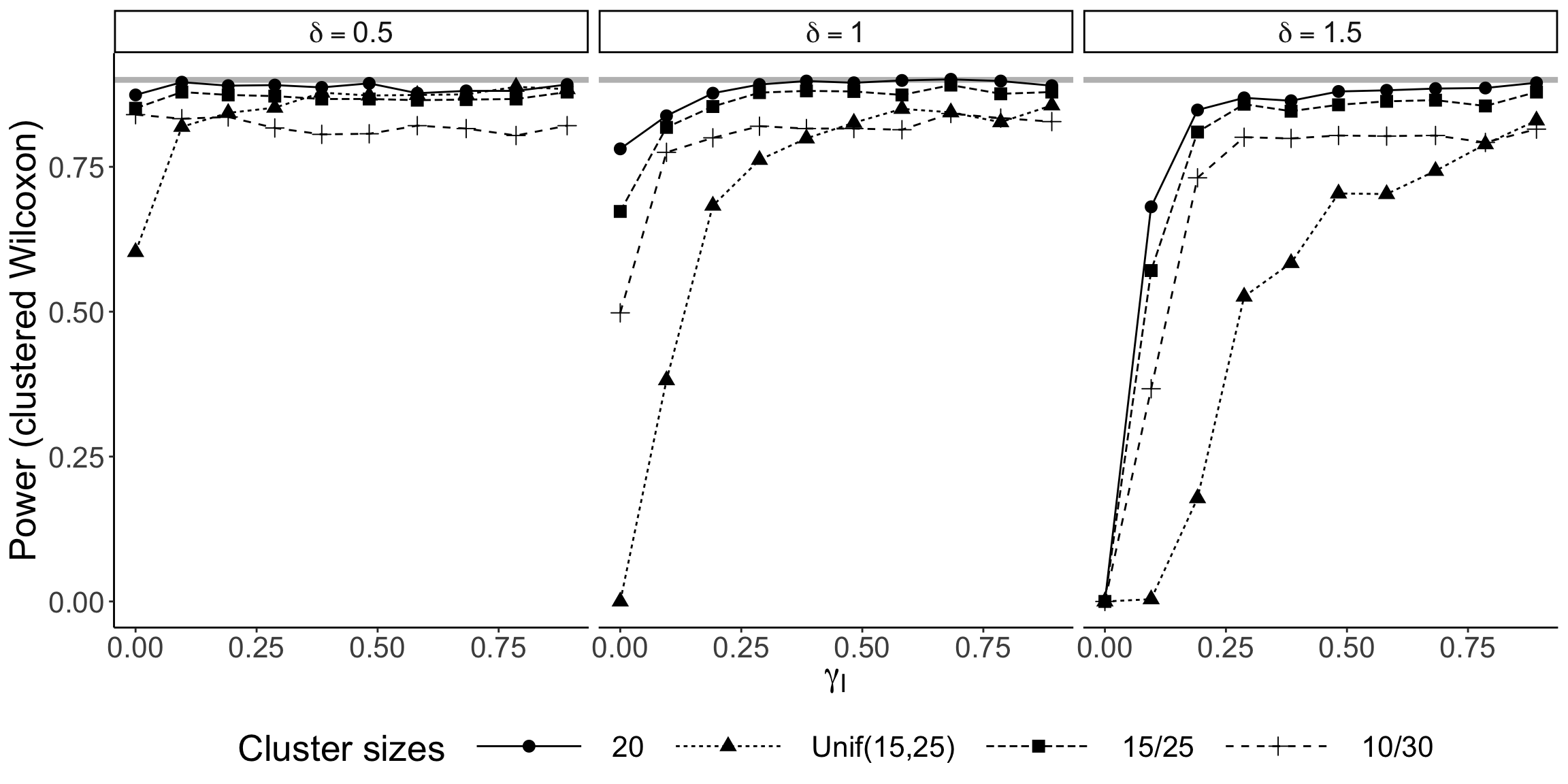}
    \caption[Power of clustered Wilcoxon rank-sum tests when actual cluster sizes differ from predetermined cluster sizes]{Number of clusters per arm calculated based on predetermined equal cluster sizes of 20, and power under equal or unequal cluster sizes of the actual sample data. The power was obtained based on clustered Wilcoxon rank-sum tests.}
    \label{fig:enter-label}
\end{figure}

\clearpage
\section{Web Figure 4}
For ordinal outcomes in individual RCTs, $X_i$ and $Y_j$ were generated by discretizing $X_{0i}$ (i.e., $X_{0i} \stackrel{i.i.d}{\sim} N(0,1)$) and $Y_{0j}$ (i.e., $Y_{0j} \stackrel{i.i.d}{\sim} N(0,\mu)$) by quantiles of a standard normal distribution (i.e., using $1/k$, $2/k$,..., $(k-1)/k$ quantiles for $k$ levels. For example, using the 1/2 quantiles for 2 levels; the 0.2, 0.4, 0.6, 0.8 quantiles for 5 levels; and the 0.1, 0.2, ..., 0.8, 0.9 quantiles for 10 levels). The value of $\theta$ can be calculated from $\delta$ \citep{de_neve2019}, where $\delta$ varies between $0.5$ and $1.5$. We then can calculate $\mu$ from $\theta$ by $\mu = \sqrt{2}\Phi^{-1}(\theta)$.  We derived $\bar{\pi}_l$ based on the known distributions and empirically computed values for $\gamma_I$ and $\delta$ based on the specific data generation scenario, and then calculated sample sizes for each ordinal variable. The power simulations were conducted 1,000 times for each value of $\delta$. 

\begin{figure}[!h]
     \centering
     \begin{subfigure}[b]{0.69\textwidth}
         \centering 
         \includegraphics[width=1\textwidth]{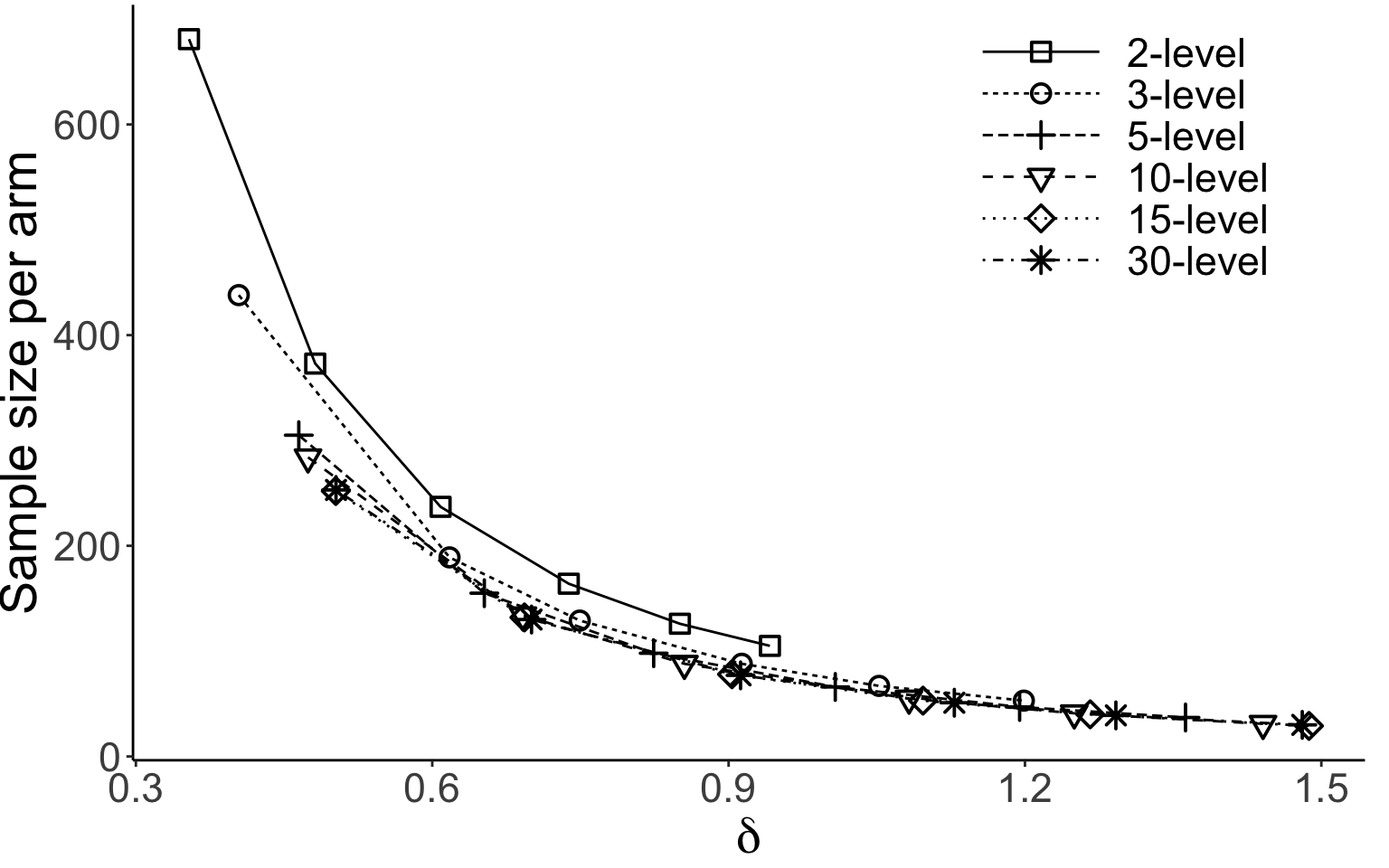}
     \end{subfigure}\\
          \begin{subfigure}[b]{0.7\textwidth}
         \centering
         \includegraphics[width=1\textwidth]{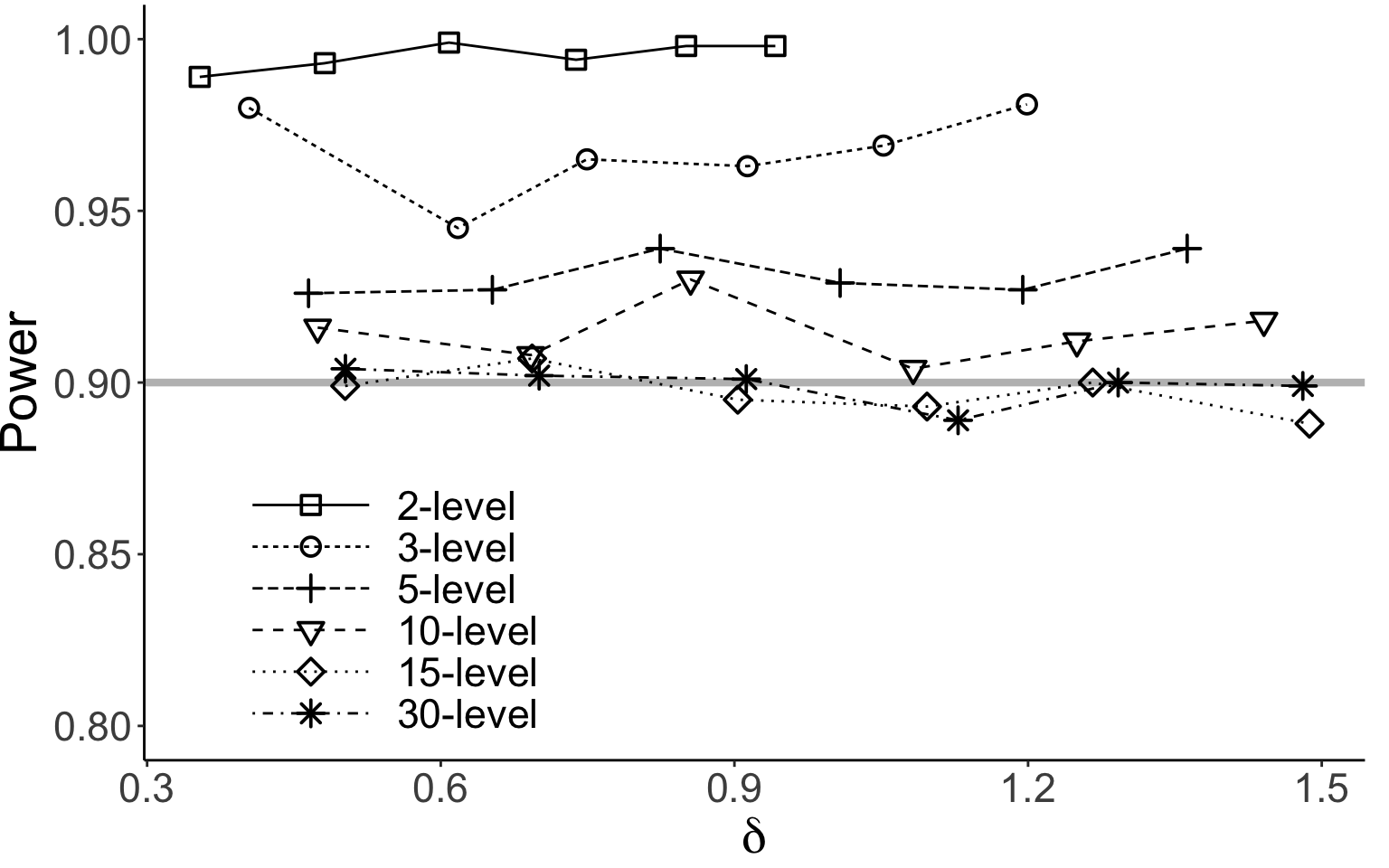}
     \end{subfigure}
     \caption[]{Sample size per arm (calculated by Whitehead's formula) and power for independent ordinal data.}
\end{figure}

\clearpage
\section{Web Figure 5}
We conducted simulations to compare Whitehead's sample size formula with two commonly used sample size formulas (unpooled \citep{chow2008} and pooled \citep{fleiss2003}) for binary outcomes, which are (per arm)  
$$n_1 = \frac{(z_{\alpha/2}+z_{\beta})^2\{p_t(1-p_t)+p_c(1-p_c)\}}{(p_t-p_c)^2},$$
$$n_2 = \frac{\big(z_{\alpha/2} \sqrt{2\bar{p}(1-\bar{p})}+z_{\beta}\sqrt{p_t(1-p_t)+p_c(1-p_c)}\big)^2}{(p_t-p_c)^2},$$
where $p_t$ and $p_c$ are the proportions of the treatment and control arms, respectively. The simulation results show that the sample sizes calculated by the three approaches for binary outcomes were nearly the same. 
\begin{figure}[!h]
    \centering
    \includegraphics[width=0.7\linewidth]{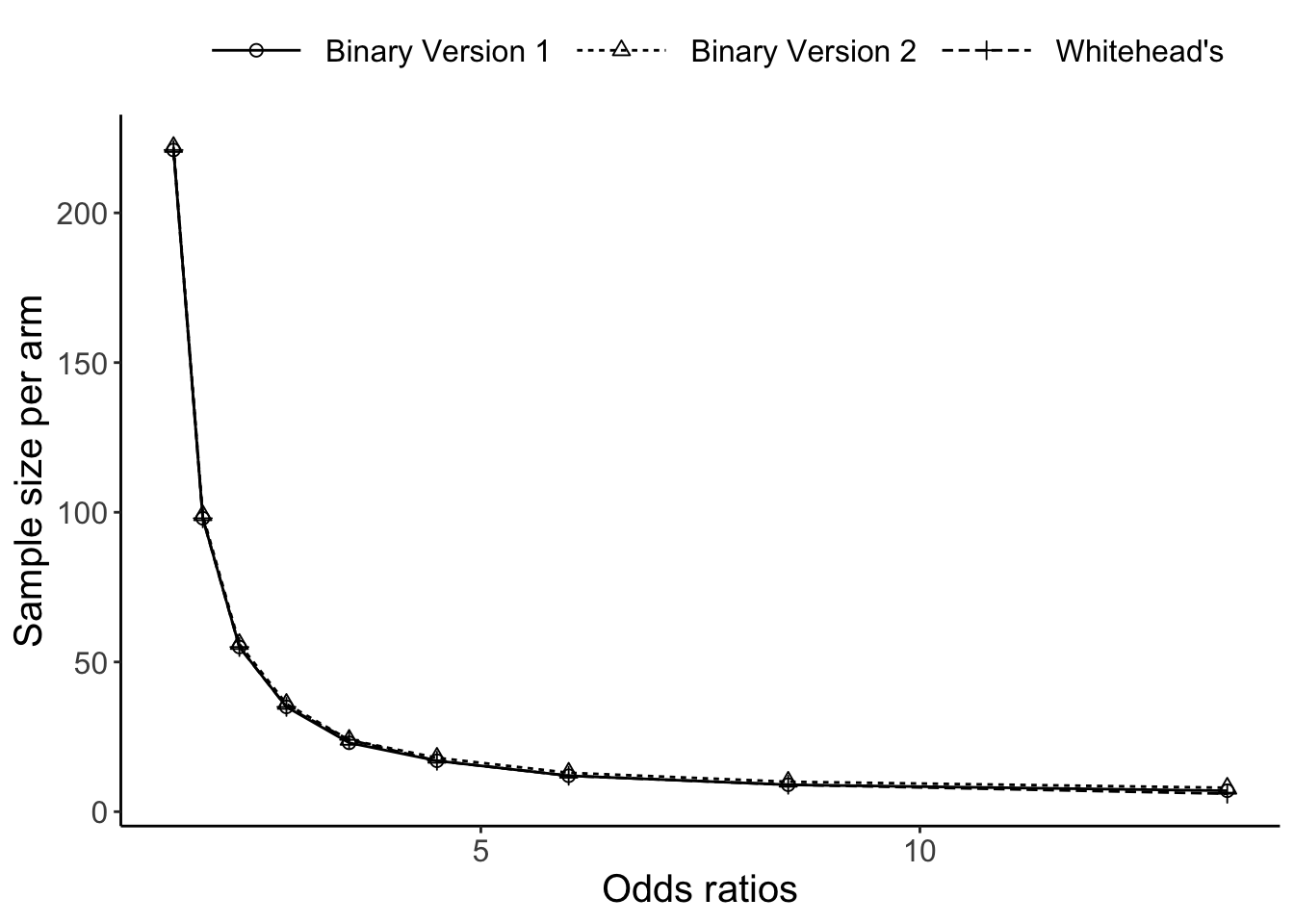}
    \caption{Sample size per arm calculated by Whitehead's sample size formula and two commonly used sample size formulas for binary outcomes}
\end{figure}

\clearpage
\section{Web Figure 6}
\begin{figure}[!h]
    \centering
    \includegraphics[width=1\textwidth]{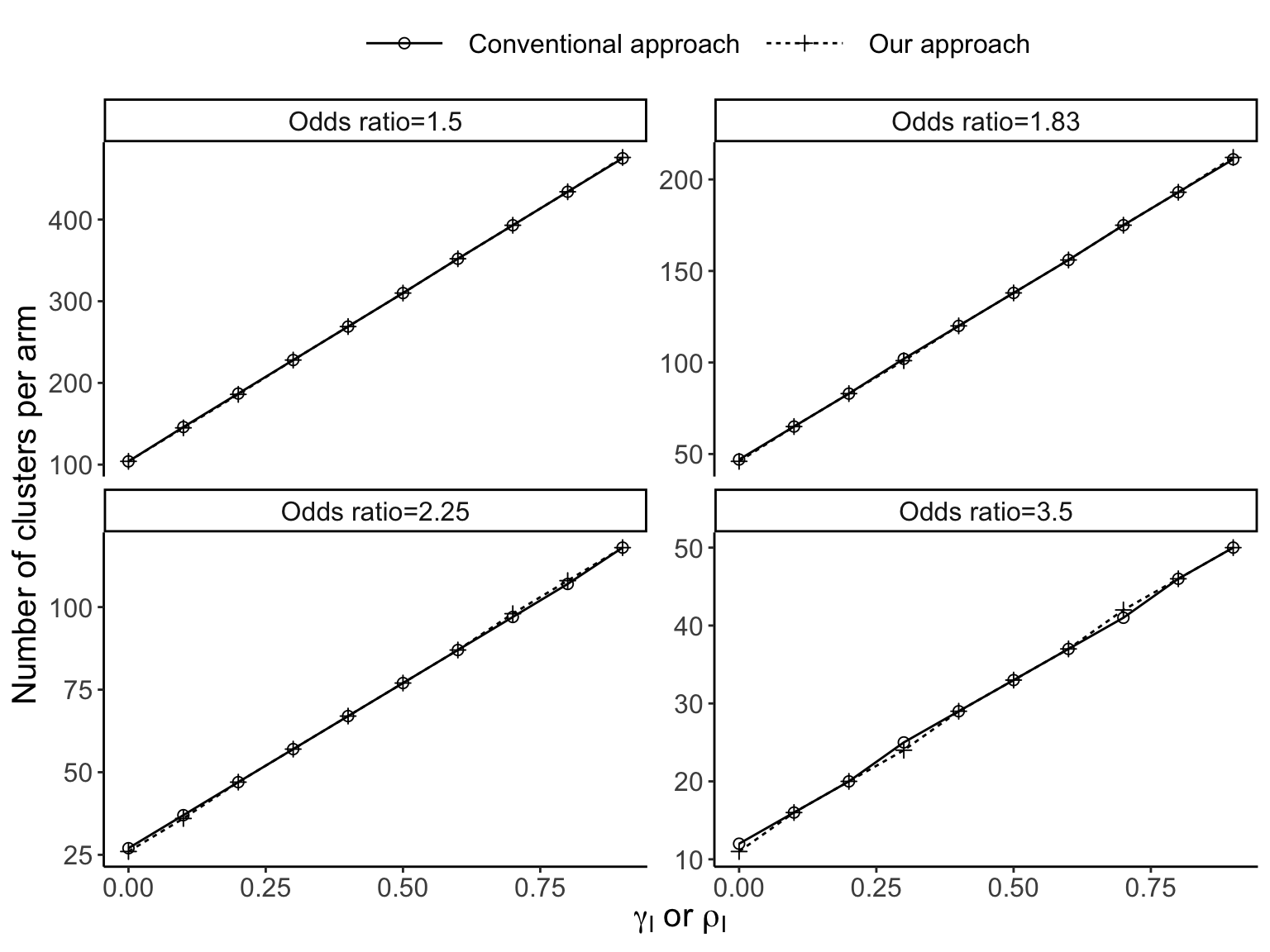}
    \caption{Number of clusters per arm calculated by the conventional approach and our approach for binary outcomes. The cluster size was 5. 
    The two-sided significance level was set to 0.05 and the power was set to 0.9. With binary outcomes, the rank ICC $\gamma_I$ is equal to the ICC $\rho_I$.}
\end{figure}

\clearpage
\bibliographystyle{biom} 
\bibliography{reference.bib}

\label{lastpage}